\documentclass[aps,prd,floatfix,nofootinbib,superscriptaddress,twocolumn,showkeys]{revtex4-2}

\usepackage{amsmath,amsfonts,extarrows,amssymb,mathrsfs,times,bm}
\usepackage{extarrows}
\usepackage{graphicx}
\usepackage{float}
\usepackage{graphicx,epstopdf}
\usepackage{mathtools}
\usepackage[colorlinks=true,breaklinks=true,linkcolor=blue,citecolor=blue,urlcolor=blue]{hyperref}
\usepackage{tikz}
\usepackage[usenames,dvipsnames]{pstricks}
\usepackage{epsfig}

\newcommand\nn{{\nonumber}}

\usepackage{allpurpose}

\usepackage{ulem}
\usepackage{color}

\newcommand{\delred}[1]{{\color{red}{\ifmmode\text{\sout{\ensuremath{#1}}}\else\sout{#1}\fi}}}

\begin{document}

\title{Deflection and gravitational lensing with finite distance effect in the strong deflection limit in stationary and axisymmetric spacetimes}

\author{Yujie Duan}
\thanks{These authors contributed equally to this work.}
\address{School of Physics and Technology, Wuhan University, Wuhan, 430072, China}

\author{Siyan Lin}
\thanks{These authors contributed equally to this work.}
\address{School of Physics and Technology, Wuhan University, Wuhan, 430072, China}

\author{Junji Jia}
\email[Corresponding author:~]{junjijia@whu.edu.cn}
\address{Department of Astronomy \& MOE Key Laboratory of Artificial Micro- and Nano-structures, School of Physics and Technology, Wuhan University, Wuhan, 430072, China}

\date{\today}

\begin{abstract}
We study the deflection and gravitational lensing (GL) of both timelike and null signals in the equatorial plane of arbitrary stationary and axisymmetric spacetimes in the strong deflection limit. Our approach employs a perturbative method to show that both the deflection angle and the total travel time take quasi-series forms $\displaystyle \sum_{n=0}\lsb C_n\ln (1-b_c/b)+D_n\rsb  (1-b_c/b)^n$, with the coefficients $C_n$ and $D_n$ incorporating the signal velocity and finite distance effect of the source and detector. This new deflection angle allows us to establish an accurate GL equation from which the apparent angles of the relativistic images and their time delays are found. These results are applied to the Kerr and the rotating Kalb-Ramond (KR) spacetimes to investigate the effect of the spacetime spin in both spacetimes, and the effective charge parameter and a transition parameter in the rotating KR spacetime on various observables. Moreover, using our approach, the effect of the signal velocity and the source angular position on these variables  is also studied.
\end{abstract}


\maketitle

\section{Introduction}

Deflection of light rays in the strong deflection limit (SDL) has drawn vast  attention in recent years, especially after the observation of black hole (BH) shadows \cite{EventHorizonTelescope:2019dse,EventHorizonTelescope:2022wkp}.
Earlier, light deflection was predominately used for gravitational lensing (GL) in the weak deflection limit, including the astronomical strong GL, the microlensing and weak lensing phenomena. These effects have since become indispensable tools in astronomy, from measuring the galaxy (cluster) mass distributions \cite{Limousin:2006cv,Halkola:2006pq}, constraining the cosmological constant \cite{DES:2018ufa,DES:2017qwj,Planck:2018vyg} and dark energy/matter parameters \cite{Clowe:2006eq,Coe:2012kj,Hezaveh:2014aoa}, to examining the gravitational theories beyond general relativity (GR) \cite{Keeton:2005jd,Joyce:2016vqv}.

For the deflection of trajectories in the SDL, its importance is most evident in the study of phenomena such as BH accretion or BH shadows \cite{Brito:2015oca, Cunha:2018acu}. Theoretically, research into the deflection and GL of light rays in the SDL began  a long time ago \cite{Darwin:1959aa,1987AmJPh..55..428O} and was revived more recently \cite{Virbhadra:1999nm, Virbhadra:2002ju,Bozza:2002zj,Bozza:2002af,Bozza:2003cp,Bozza:2010xqn}. To date, deflection and GL in the SDL have been investigated in numerous BH spacetimes  \cite{Wei:2011nj,Hsieh:2021scb,Wang:2016paq,Kuang:2022xjp,Chen:2016hil,Islam:2021dyk,Islam:2021ful,Hsieh:2021rru} and spacetimes of other more exotic types such as wormholes \cite{Tsukamoto:2016qro}, naked singularities \cite{Gyulchev:2008ff}, or compact objects \cite{Chen:2012kn} etc.

Traditionally, the messengers experiencing these deflections or lensings have always been photons. However, with the development of cosmic ray (CR) physics (see \cite{AMS:2021nhj} and references therein), the discovery of cosmological neutrinos (SN 1987A
\cite{Bionta:1987qt,Kamiokande-II:1987idp}, blazar TXS 0506 + 056 \cite{IceCube:2018dnn,IceCube:2018cha}, galaxy NGC 1068 \cite{IceCube:2022der}, high energy neutrino spectrum \cite{Stettner:2019tok}) and the detection of gravitational waves \cite{LIGOScientific:2016aoc,LIGOScientific:2017vwq} in recent years, it becomes very clear that these particles can also act as messengers from the source to our detector.
When they pass by a massive compact object, they will also be strongly deflected. Among these signals, CR and neutrinos are known to have nonzero mass and therefore will travel along timelike geodesics, unlike photons or gravitational waves in GR.

Recently, we have investigated the deflection and GL of such particles, as well as those of null signals, in the SDL in static and spherically symmetric spacetimes \cite{Jia:2020qzt,Liu:2021ckg,Zhou:2022dze}. Our approach employs a perturbative method that automatically accounts for the finite distance effect of the source and detector. One advantage of this method over some other works that used only deflection angle for  the source and detector at infinity, is that it allows us to establish a GL equation that can deal very naturally with these finite distance effects and the case that the source is not well aligned with the lens-observer axis \cite{Jia:2020qzt}. In this work, we will study the deflection and GL of both timelike and null signals in the SDL in arbitrary stationary and axisymmetric (SAS) spacetimes. Note that the study of the finite distance effect to the deflection angles and/or GL for null signals using the Gauss-Bonnet theorem method was initially carried out in Ref. \cite{Ishihara:2016vdc} in the weak deflection limit and then generalized to the SDL \cite{Ishihara:2016sfv}, the SAS  \cite{Ono:2017pie} and 
asymptotically nonflat spacetimes \cite{Takizawa:2020egm}, the timelike signal case \cite{Li:2019qyb} as well as with electrostatic interaction \cite{Li:2020ozr,Li:2021xhy}. 

Our motivation for this work is threefold. Firstly, we would like to see the effect of spacetime spin and other spacetime parameters on the deflection and GL of subluminal signals in the SDL. Such signals might experience extra deflection depending on their motion direction and deviation of velocity from light. Secondly, we aim to test whether our SDL perturbative method can be generalized smoothly to arbitrary SAS spacetimes by studying the deflection and GL in both Kerr and some other SAS spacetimes.
We show that our previously developed method can be safely applied to arbitrary SAS spacetimes for signals with arbitrary velocity. Moreover, the result of the deflection or total travel time still takes  the form of
\begin{align}
\Delta\phi~\text{or}~\Delta t=\sum_{n=0}^\infty \lsb C_n\ln\lb1-\frac{b_c}{b}\rb +D_n\rsb \lb1-\frac{b_c}{b}\rb^n,\nn
\end{align}
and our method can be used to compute the coefficients to high orders. These coefficients incorporate the information of various spacetime and signal parameters, as well as the finite distance effect. Deflection angle with finite distance effect naturally enables us to find the apparent angles for arbitrary source angular locations.

As will be shown in this work, the perturbative method we developed previously for static and spherically symmetric spacetimes in the SDL can be well generalized to {\it arbitrary} SAS spacetimes, with the results expressed as functions of the SAS metric function parameters. This implies that essentially the deflection and travel time in all SAS spacetimes with known metrics in the SDL are readily known. Moreover, the results as given in Eqs. \eqref{eq:sedphi} and \eqref{eq:sedt} not only contain the finite distance effect, but also the higher order terms, and work for timelike signals in addition to light rays. With the development of the Gauss-Bonnet  theorem-based methods \cite{Ishihara:2016vdc, Ishihara:2016sfv, Ono:2017pie, Takizawa:2020egm,Li:2019qyb,Li:2020ozr,Li:2021xhy}, the computation of the deflection angles including in the SDL and with finite distance effect has also advanced tremendously. However our method provides a special and systematical way to tackle the  time-related quantities in the SDL, on which there have been far fewer works so far. 

The paper is organized as follows.
In Sec. \ref{sec:pm}, we lay out the preliminaries, discuss details of  the perturbation method and obtain the general results for the deflection angle and total travel time in arbitrary SAS spacetimes. In Sec. \ref{sec:gl}, we present the GL equation and solve for the apparent angles of relativistic images and the time delay between them. In Sec. \ref{sec:apptoss}, we apply the results obtained in previous sections to two exemplary SAS spacetimes: the Kerr and the rotating Kalb-Ramond (KR) spacetime. We will show the effect of various parameters, including the spacetime spin, signal velocity and the effective charge and transition parameters of the rotating KR spacetime, on the deflection angle, the apparent angles and the time delay of the images.
In Sec. \ref{sec:disc}, we conclude the paper with a brief summary and discussion.

\section{The perturbative method \label{sec:pm}}

We start from the most general line element of  an SAS spacetime. In the Boyer-Lindquist coordinates, it can always be expressed as
\begin{align}
\dd s^2 = -A\dd t^2 + B \dd t \dd \phi + C \dd \phi^2 + D \dd r^2 + F \dd \theta^2,
\end{align}
where $t,~r,~\theta,~\phi$ are coordinates and $A,~B,~C,~D,~F$ are functions of $r$ and $\theta$ only. For simplicity, we restrict our attention to the geodesic in the equatorial plane. Setting $\theta=\pi/2$, the above line element becomes
\begin{align}\label{eq:spacetime2}
\dd s^2 = -A(r)\dd t^2 + B(r) \dd t \dd \phi + C(r) \dd \phi^2 + D(r) \dd r^2 ,
\end{align}
where we have suppressed the dependence of the metric functions on $\theta$. With this metric, it is easy to obtain the geodesic equations
\begin{align}
\dot{\phi}&=\frac{2 (2 L A-E B)}{B^2+4 A C}, \label{eq:phideq}\\
\dot{t}&= \frac{2 (L B+2 E C)}{B^2+4 A C}, \label{eq:tdeq}\\
\dot{r}^2 &=\frac{( E^2 - \kappa A) (B^2+4AC) - (2LA-E B)^2}{AD \left(B^2+4 A C\right)}, \label{eq:rdeq}
\end{align}
where dot stands for derivative with respect to the proper time of a timelike signal or the affine parameter of a null signal. Here, $L$ and $E$ are the angular momentum and energy (per unit mass) of the signal respectively, and $\kappa=0$ and $1$ for null and timelike particles respectively.
$L$ and $E$ can be related to the impact parameter $b$ and asymptotic velocity $v$ of the timelike signal via
\begin{align}
E&=\frac{1}{\sqrt{1-v^2}},\label{eq:erelv}\\
|L|&=\left|(\textbf{v}\times \textbf{r})\cdot \hat{\textbf{z}}\right|\Big|_{r\to\infty}=\frac{b v}{\sqrt{1-v^2}}\label{eq:lrelv}\\
&=b\sqrt{E^2-\kappa},~(b>0) \label{eq:lerela}
\end{align}
where relation \eqref{eq:lerela} works for null signals too.

Using Eqs. \eqref{eq:phideq}-\eqref{eq:rdeq}, we can find
\begin{align}
        &\frac{\dd \phi}{\dd r}=\frac{\sqrt{AD}( 4LA-2EB)}{\sqrt{B^2+4 A C }} \nn\\
        &\times\frac{1}{\sqrt{( E^2 - \kappa A ) ( B^2+4AC) - (2LA-E B)^2}},\label{eq:dphdr}\\
        &\frac{\dd t}{\dd r}=\frac{\sqrt{AD}( 2LB+4EC)}{\sqrt{B^2+4 A C }} \nn\\
        &\times\frac{1}{\sqrt{( E^2 - \kappa A ) ( B^2+4AC) - (2LA-E B)^2}}. \label{eq:dtdr}
\end{align}
Using these equations, the deflection angle and the total travel time of a signal originating from a source at $r_s$ to a detector at $r_d$ can be expressed as
\begin{align}
    \Delta\phi=\left( \int_{r_0}^{r_s}+\int_{r_0}^{r_d}\right)\frac{\dd \phi}{\dd r}\dd r,\label{eq:dphidef}\\
    \Delta t=\left( \int_{r_0}^{r_s}+\int_{r_0}^{r_d}\right)\frac{\dd t}{\dd r}\dd r,\label{eq:dtdef}
\end{align}
where $r_0$ is the radius of the pericenter of the geodesic defined by
\begin{align}\label{eq:dr=0}
    \dot{r}|_{r=r_0}=0.
\end{align}
Substituting Eq. \eqref{eq:rdeq} into above, we can relate $r_0$ to the angular momentum $L$ of the signal, i.e.,
\begin{align}\label{eq:L}
    L=\frac{E B(r_0) +s_1 \sqrt{(E^2 - \kappa A(r_0)) ( B^2(r_0)+4A(r_0)C(r_0))} }{2 A(r_0)}.
\end{align}
Here $s_1=\pm 1$ is the sign introduced when taking the square root of Eq. \eqref{eq:rdeq}. $s_1=+1$ and $-1$ correspond to the anti-clockwise and clockwise direction of rotation of the trajectory at $r=r_0$ respectively.
Using Eq. \eqref{eq:lerela} in the above, a local  correspondence between $b$ and $r_0$ can be established
\begin{align}
\frac{1}{b}=\frac{|2 v A(r_0) |}{| B(r_0)+s_1 H(r_0)|},
\label{eq:bctwovalues}
\end{align}
where we have defined
\begin{align}
    H(r)=\sqrt{\left[1-(1-v^2)A(r) \right](B(r)^2+ 4A(r)C(r))}. \label{eq:hrdef}
\end{align}
For convenience in later computation, we can also drop the absolute sign by adding an $s_2=\mathrm{sign}(L)=\pm1$ so that this becomes
\begin{align}
\frac{1}{b}=\frac{2 s_2 v A(r_0)}{ B(r_0)+s_1 H(r_0)}.
\label{eq:1b_r0}
\end{align}
Eq. \eqref{eq:bctwovalues} suggests that for each $r_0$ there are effectively two impact parameters corresponding to the two signs of $s_1$.

For black hole spacetimes, if we adjust the approach direction of the trajectory so that $r_0$ decreases, it will reach a critical value $r_c$ below which the signal will eventually enter the BH and not emerge. The value of $r_c$ therefore should satisfy not only Eq. \eqref{eq:dr=0},
but also the following condition
\begin{align}
    \frac{\dd\dot{r}^2}{\dd r} \Big|_{r=r_c}&=0. \label{eq:definerc}
\end{align}
After substituting Eq. \eqref{eq:rdeq}, Eqs. \eqref{eq:dr=0} and \eqref{eq:definerc} will allow us to solve for $r_c$ for any given metric. We note that there is no condition to constrain the number of physical solutions to these equations to only one. In other words, multiple $r_c$'s can exist for one spacetime. For example, for photons in Kerr spacetime, we can obtain an analytical expression for its two $r_c$'s \cite{Chandrasekhar:1985kt}
\begin{align}
    r_{c\mathrm{K}\gamma}=2M\left\{1+\cos\left[\frac{2}{3}\arccos\left(\frac{\pm \hat{a}}{M}\right)\right]\right\} \label{eq:rckerrk0}
\end{align}
(see the red and blue curves in Fig. \ref{fig:rcbckerr} (a).)
Corresponding to each $r_c$, Eq. \eqref{eq:1b_r0} determines a critical impact parameter $b_c$:
\begin{align}
b_c=\frac{B(r_c)+s_1 H(r_c)}{2s_2 v A(r_c)}.
\label{eq:1b_rc}
\end{align}
Note that although in the weak deflection limit for each large $r_0$ in Eq. \eqref{eq:bctwovalues} there could be two different $b$'s, this is usually not true for $r_c$ and $b_c$, which are in the SDL and much smaller. The mathematical reason for this is the appearance of the additional critical condition \eqref{eq:definerc}. In principle, it can  be proven that if $r_c$ depends nontrivially on $\mathrm{sign}(L)$, there will be only one $b_c$ for each $r_c$. This implies that in this case, once $r_c$ is fixed, the signs $s_1$ and $s_2$ are also fixed.
For example, for photons in Kerr spacetime, corresponding to the two branches of $r_c$'s in Eq. \eqref{eq:rckerrk0} there are also only two branches of $b_c$'s \cite{Chandrasekhar:1985kt}
\begin{align}
    b_{c\mathrm{K}}=\pm \hat{a}+6M\cos\left[\frac{1}{3}\arccos\left(\frac{\pm\hat{a}}{M}\right)\right]
\end{align}
(see the red curves in Fig. \ref{fig:rcbckerr} (b)). Furthermore however, without specifying the explicit form of the metric functions, it is not possible to determine {\it a priori} what choice $s_1$ and $s_2$ will take for a given $r_c$. Therefore we will retain these signs in $b_c$ as given in Eq. \eqref{eq:1b_rc} until we apply it in specific spacetimes in Sec. \ref{sec:apptoss}.

One of our main objectives in this work is then to compute the deflection angle \eqref{eq:dphidef} and travel time \eqref{eq:dtdef} of signals in the SDL, i.e., when $b$ approaches $b_c$ or equivalently, when $r_0$ approaches $r_c$. However, since these integrals cannot be carried out analytically for most spacetimes, some approximation is usually needed.
Here we will generalize a change of variables introduced in Ref. \cite{Jia:2020qzt} and use it to carry out a perturbative calculation. To this end, we will think of $b$ in Eq. \eqref{eq:1b_r0} as a function of $r_0$ and first define a function $p(1/r_0)$ using it as well as the constant $1/b_c$ in Eq. \eqref{eq:1b_rc}, as
\begin{align}
p\left( \frac{1}{r_0}\right) =&\frac{1}{b_c}- \frac{1}{b(r_0)}
\equiv\frac{a}{b_c}. \label{eq:pp}
\end{align}
In the last step, we defined  $a\equiv 1-b_c/b(r_0)$, which characterizes how close the trajectory is from the critical case.
Changing $r_0$ to $r$, this can also be written as
\begin{align}
p\left(\frac{1}{r}\right)=\frac{1}{b_c}- \frac{1}{b(r)}=\frac{\xi}{b_c}.\label{eq:pxi} \end{align}
Similarly, in the last step, we defined a new variable $\xi \equiv  1-b_c/b(r)$, which characterizes how close a point with radius $r$ is to the critical point.  In this work we explicitly concentrate on the case that $\xi>0$ to make sure that the source and observer are outside the critical radius $r_c$. Clearly, when $r=r_0$, we have $\xi=a$.
Denoting the inverse function of $p(x)$ as $q(x)$, from Eq. \eqref{eq:pxi}, we have
\begin{align}
\frac{1}{r} = q\left(\frac{\xi}{b_c} \right) \label{eq:q1r} .\end{align}
We emphasize that once the metric functions of the spacetime are known, the functions $p(x)$ can be immediately determined. Although inverting a function is not always possible, fortunately what is needed here is the perturbative form of $q(x)$ in the SDL $\xi\to0$ and this can always be obtained using the  Lagrange inversion theorem.

Using Eq. \eqref{eq:q1r} as a change of variables from $r$ to $\xi$ and Eq. \eqref{eq:lrelv} to change $L$ to $b$ and $v$, the terms in Eq. \eqref{eq:dphidef} and Eq. \eqref{eq:dtdef} can be transformed to
\begin{subequations} \label{eq:varinttfrtoxi}
\begin{align}
 &r \to \frac{1}{q},\\
 &L \to s_2 b\frac{v}{\sqrt{1-v^2}},\\
 & r_0 \to 1-\frac{b_c}{b}= a,~r_i \to b_c\; p\lb  \frac{1}{r_i}\rb \equiv \eta_i ~(i=s,d), \label{eqone}  \\
 & \dd r \to -\frac{q'}{q^2} \frac{\dd \xi}{b_c},
\end{align}
\end{subequations}
where $q$ and $q'$ stand for $q(\xi/b_c)$ and $q'(\xi/b_c)$ respectively. Subsequently, $\Delta\phi$ and $\Delta t$ become respectively
\begin{align}
    \Delta\phi&=\left( \int_{a}^{\eta_s}+\int_{a}^{\eta_d}\right)\frac{y_{\phi}(\xi)}{\sqrt{\xi-a}(\sqrt{\xi}+\zeta_h)}\dd \xi,\label{eq:dphitrans}\\
    \Delta t&=\left( \int_{a}^{\eta_s}+\int_{a}^{\eta_d}\right)\frac{y_t(\xi)}{\sqrt{\xi-a}(\sqrt{\xi}+\zeta_h)}\dd \xi,\label{eq:dttrans}
\end{align}
where
\begin{align}
&y_{\phi}(\xi)=\frac{\sqrt{AD}}{\sqrt{B^2+4AC}}\frac{q'}{b_c q^2}(\xi-1)(\sqrt{\xi}+\zeta_h)\nn\\
		&\times\frac{2s_2 v -(1-a)B/(b_cA)}{\sqrt{(2-a-\xi)v^2+s_2 B(\xi-1)(1-a)v/(b_c A)}},\label{eq:yphidef}\\
&y_t(\xi)=\frac{s_2 b_c v B+2(1-a)C}{2s_2 b_c v A-(1-a)B}y_{\phi}(\xi), \label{eq:ytdef}
\end{align}
and all $A,~B,~C,~D$ are evaluated at $1/q$. Here
\be \zeta_h=\sqrt{b_c\,p(1/r_h)}=\sqrt{1- b_c/b(r_h)}
\ee
with $r_h$ being the radius of the outermost horizon of the spacetime which is determined by $B(r_h)^2+4A(r_h)C(r_h)=0$~\cite{Konoplya:2021slg}.

To proceed, we will expand these integrands in the SDL $\xi\to 0$ and calculate these integrals perturbatively. Expanding the  $y_{\phi}(\xi)$ in Eq. \eqref{eq:dphitrans} for small $\xi$ first, we find that its series always takes the form
\begin{align}
        y_{\phi}(\xi)=\sum_{n=-1}^{\infty}y_{\phi,n}(a)\; \xi^{\frac{n}{2}}, \label{eq:yphindef}
\end{align}
where $y_{\phi,n}(a)$ are the expansion coefficients determined by the metric functions and the location of $r_c$. We will show how to calculate them and their exact form at the end of this section. The crucial point to note is that these $y_{\phi,n}(a)$ might depend on $a$ but not on $\xi$, and therefore, will not complicate the integration over $\xi$.
Now expanding the first factor in Eq. \eqref{eq:ytdef} as
\begin{align}\label{eq:gn}
 \frac{s_2 b_c v B+2(1-a)C}{2s_2 b_c v A-(1-a)B}=\sum_{n=0}^{\infty}g_n(a)\; \xi^{\frac{n}{2}},
\end{align}
then we see that  $y_{t}(\xi)$ can also be expanded into the form
\begin{align}
        y_{t}(\xi)&=\sum_{n=0}^{\infty}g_n(a) \;\xi^{\frac{n}{2}}\sum_{n=-1}^{\infty} y_{\phi,n}(a) \;\xi^{\frac{n}{2}}\nn\\
        &\equiv\sum_{n=-1}^{\infty}y_{t,n}(a)
        \;\xi^{\frac{n}{2}}, \label{eq:ytndef}
\end{align}
where in the last step we combined the coefficients $y_{\phi,n}(a)$ and $g_n(a)$ into $y_{t,n}(a)$.

Substituting Eqs. \eqref{eq:yphindef} and \eqref{eq:ytndef} into Eq. \eqref{eq:dphitrans} and Eq. \eqref{eq:dttrans} respectively, we observe that both $\Delta\phi$ and $\Delta t$ become series of integrals
\begin{align}
    \Delta\phi&=\sum_{n=-1}^\infty y_{\phi,n} (a)\sum_{i=s,d} \int_{a}^{\eta_i}\frac{\xi^{\frac{n}{2}}}{\sqrt{\xi-a}(\sqrt{\xi}+\zeta_h)}\dd \xi,\label{eq:dphitransexp}\\
    \Delta t&=\sum_{n=-1}^\infty y_{t,n} (a)\sum_{i=s,d} \int_{a}^{\eta_i}\frac{\xi^{\frac{n}{2}}}{\sqrt{\xi-a}(\sqrt{\xi}+\zeta_h)}\dd \xi.\label{eq:dttransexp}
\end{align}
This kind of integrals can always be worked out as shown in detail in Appendix \ref{appendix:integrals}. Since we are in the SDL, we can expand the integration results in the $a\to0$ limit. Therefore, directly using Eqs. \eqref{eq:imoneexp} and \eqref{eq:ilargernexp}, the $\Delta\phi$ and $\Delta t$
are found to take quasi-series forms of small $a$
\begin{align}
&\Delta\phi=\sum^\infty_{n=0}[C_{\phi,n}\ln{a}+D_{\phi,n}]a^{n},
    \label{eq:sedphi}\\
&\Delta t=\sum^\infty_{n=0}[C_{t,n}\ln{a}+D_{t,n}]a^{n},
    \label{eq:sedt}
\end{align}
where the first set of parameters, $C_{\alpha,0}$ and $D_{\alpha,0}~(\alpha=\phi,t)$, are
\begin{align}
    &C_{\alpha,0}=-\frac{2y_{\alpha,-1}(0)}{\zeta_h},\label{eq:calpha0}\\
    &D_{\alpha,0}=\sum_{i=s,d}\left\{\frac{2y_{\alpha,-1}(0)}{\zeta_h}\ln{\left(\frac{2\zeta_h\sqrt{\eta_i}}{\zeta_h+\sqrt{\eta_i}}\right)}\right.\nn\\
    &+\sum_{n=0}^\infty2y_{\alpha,n}(0)\left[(-\zeta_h)^n\ln{\left(\frac{\sqrt{\eta_i}+\zeta_h}{\zeta_h}\right)}\right.\nn\\
    &\left.\left.+\sum_{j=1}^nC_n^j\frac{(-\zeta_h)^{n-j}}{j}\left[(\sqrt{\eta_i}+\zeta_h)^j-\zeta_h^j\right]\right]\right\}.
    \label{eq:cdfirstfew}
\end{align}
When $a$ is small enough, we can show later in Sec. \ref{sec:gl} that the formula \eqref{eq:sedphi} for the deflection angle will be accurate enough even at the first order. 

 We note from Eq. \eqref{eq:calpha0} that the coefficients $C_{\alpha,0}$ do not depend on $\eta_{s,d}$ and therefore they will not change as the finite distance of the source/detector varies. On the other hand, for coefficients $D_{\alpha,0}$ in Eq. \eqref{eq:cdfirstfew},  their dependence on the finite distance of $r_{s,d}$ is explicit. 
We can argue that these $D_{\alpha,0}$ should decrease as the source/detector radii $r_{s,d}$ decrease, assuming the impact parameter for the signal remains unchanged. The reason is that as $r_{s,d}$ decrease, both $\Delta \phi$ and $\Delta t$ become smaller (refer to Fig. \ref{fig:schm}), while the $C_{\alpha,0}\ln a$ term remains unchanged when $b$ and consequently $a$ is not changed. Therefore, the only reason for the decrease of $\Delta\phi$ and $\Delta t$ is the decrease of $D_{\alpha,0}$.

\subsection*{Computation of $y_{\phi,n}(0)$ and $y_{t,n}(0)$}

In the above procedure to obtain $\Delta \phi$ and $\Delta t$ as given in Eqs. \eqref{eq:sedphi} and \eqref{eq:sedt}, one important step we
skipped is the computation of the coefficients $y_{\phi,n}(0)$ in Eq. \eqref{eq:yphindef} and $y_{t,n}(0)$ in Eq. \eqref{eq:ytndef}. Here we present in detail how to compute them.

We note that since all terms in Eq. \eqref{eq:dphitrans} are completely fixed once the metric functions are known, these expansion coefficients will also be fixed by the metric functions.
In the SDL of $r\to r_c$, let us assume that the metric functions in Eq. (\ref{eq:spacetime2}) can be expanded into the following series
\begin{subequations}\label{eq:gbyrc}
\begin{align}
A(r\to r_c)=& \sum_{n=0}^{\infty} a_n (r-r_c)^n, \\
B(r\to r_c)=& \sum_{n=0}^{\infty} b_n (r-r_c)^n, \\
C(r\to r_c)=& \sum_{n=0}^{\infty} c_n (r-r_c)^n, \\
D(r\to r_c)=& \sum_{n=0}^{\infty} d_n (r-r_c)^n,
\end{align}
\label{eq:metricexpansion}
\end{subequations}
then substituting them into Eq. \eqref{eq:1b_rc}, we first obtain the formula for $b_c$ in terms of coefficients $a_0,~b_0,~c_0$ and $d_0$ in Eq. \eqref{eq:gbyrc}
\begin{align}
b_{c}&=\frac{b_0+s_1 H_c}{2s_2  a_0 v}=\left|\frac{b_0+s_1 H_c}{2 a_0 v}\right|,
\label{eq:bc with respect to coefficients}
\end{align}
where $H_c\equiv H(r_c)=\sqrt{\left[1-(1-v^2)a_0\right](b_0^2+4a_0c_0)}$.

Substituting Eq. \eqref{eq:gbyrc} into Eq. \eqref{eq:pp}, then the function $p(x)$ and consequently $q(x)$ can also be obtained. However, their forms for general metrics are lengthy and therefore only given in Eqs. \eqref{eq:pxseries} and \eqref{eq:qxseries} in Appendix \ref{sec:appd2} respectively. Substituting them into Eq.
\eqref{eq:yphidef} and then series expanding it, the first few $y_{\phi,n}(0)$ are obtained as
\begin{subequations}\label{eq:yphinfirstfew}
\begin{align}
& y_{\phi,-1}(0)=-\frac{s_2 r_c^2 \zeta_h q_1}{2b_c} \sqrt{\frac{s_1 H_c}{s_2 v}}\sqrt{\frac{d_{0}}{4a_{0}c_{0}+b_{0}^{2}}},\\
& y_{\phi,0}(0)=\frac{y_{\phi,-1}}{\zeta_h}+\frac{s_2  r_c^4 q_1^2\zeta_h}{b_c }\sqrt{\frac{s_1 H_c}{s_2 b_c v}}\sqrt{\frac{ d_0}{4 a_0 c_0+b_0^2}} \nn\\
&\times \left[\frac{1}{r_c}
   -\frac{q_2}{r_c^2 q_1^2}+\frac{d_1}{4d_0}+\frac{b_0 a_1-a_0b_1}{4s_1 a_0 H_c}+\frac{  b_0^2  a_1}{4 a_0 \left(4 a_0 c_0+b_0^2\right)  }\right.\nn\\
   &-\left.\frac{  2 a_0 c_1
   + b_0 b_1}{2  \left(4 a_0 c_0+b_0^2\right)}\right],
\end{align}
\end{subequations}
where $q_1$ and $q_2$ are given in Eq. \eqref{eq:q1q2}

In order to calculate $y_{t,n}(0)$, according to Eq. \eqref{eq:ytndef},
we will need to get $g_n(0)$ first. So substituting Eqs.
\eqref{eq:gbyrc} and \eqref{eq:bc with respect to coefficients} into Eq. \eqref{eq:gn}, we can obtain the first few $g_n(0)$ as
\begin{subequations}\label{eq:gnfirstfew}
\begin{align}
   g_{0}(0)=& \frac{b_0}{2 a_0}+\frac{1}{2s_1 a_0 H_c}\left(4 a_0 c_0+b_0^2 \right),\label{eq:g0exp}\\
   g_{1}(0)=&\frac{2 q_1 r_c^2 }{\sqrt{b_c} H_c^2}\left[b_0 c_1-c_0b_1-2s_2b_c v(a_0c_1-c_0a_1)\right.\nn\\
   &\left.-b_c^2v^2(a_0b_1-b_0a_1)\right]\label{eq:g1exp}.
\end{align}
\end{subequations}
Comparing Eq. \eqref{eq:g0exp} with Eq. \eqref{eq:bc with respect to coefficients}, it can be shown that for null rays $g_0(0)=s_2b_c$. Using $a_1$ solved by substituting Eq. \eqref{eq:metricexpansion} into Eqs. \eqref{eq:dr=0} and \eqref{eq:definerc}, we can also show that for null rays Eq. \eqref{eq:g1exp} yields $g_1(0)=0$.
Using Eq. \eqref{eq:ytndef}, we can express $y_{t,n}(0)$ in terms of $y_{\phi,n}(0)$ and $g_n(0)$ explicitly
\begin{align}
    y_{t,n}(0)=\sum_{m=-1}^n{y_{\phi,m}(0)g_{n-m}(0)}.\label{eq:ytningn}
\end{align}
The first few of them are \begin{subequations}\label{eq:ytnfirstfew}
\begin{align}
    y_{t,-1}(0)&=y_{\phi,-1}(0)\cdot g_{0}(0),\label{eq:ytinyphi}\\
    y_{t,0}(0)&=g_{0}(0)\cdot y_{\phi,0}(0)+g_{1}(0)\cdot y_{\phi,-1}(0).
\end{align}
\end{subequations}
Note that higher order terms of the above results \eqref{eq:yphinfirstfew}, \eqref{eq:gnfirstfew} and \eqref{eq:ytnfirstfew} can all be computed using a symbolic computation package but are too long to write out here.

\section{GL equation and solutions\label{sec:gl}}

\begin{figure}[htp!]
\centering
\includegraphics[width=0.45\textwidth]{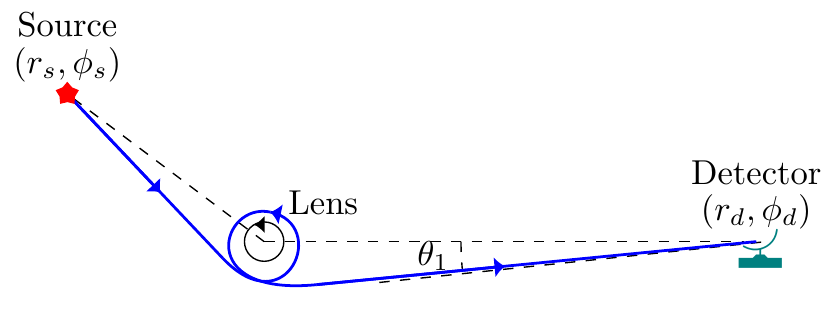}
\caption{\label{fig:schm} The deflection and lensing in the SDL. The source is located at $(r_s,~\phi_s)$ while the detector at $(r_d,~\phi_d)$. The lens has a nonzero spin and the signal is prograde in this schematic plot. }
\end{figure}

The deflection angle given by Eq. \eqref{eq:sedphi} naturally takes account into the finite distance effect of the source and detector. Then given the location $(r_s,\phi_s)$ of the static source and $(r_d,\phi_d)$ of the static detector (see Fig. \ref{fig:schm}), and the spacetime parameters $\hat{a},~M$, as well as the signal parameters $v$ or $E$, we can use it to establish a precise GL equation
\begin{align}
    \Delta\phi-(2n+1)s_1\pi=\pi-\Delta\phi_{sd}.\label{eq:exactgleq}
\end{align}
From this, we can solve for the impact parameters $b$ that allow the signal to reach the detector. This solution process has been done in Ref. \cite{Jia:2020qzt}  and the result is (after adopting the convention here)
\begin{align}
\displaystyle    b_{n}(s_1,s_2)=&\frac{b_c(s_1,s_2)}{1-\exp\lcb \frac{ \lsb (2n+1)s_1+1\rsb \pi-\Delta \phi_{sd}- D_{\phi,0}}{C_{\phi,0}}\rcb },\nn\\
&(s_1=\pm1,~n=1,2,\cdots)
    \label{eq:bnsol}
\end{align}
where $\Delta\phi_{sd}\equiv \phi_s-\phi_d=\phi_s$ and without losing any generality we have set $\phi_d=0$ as illustrated in Fig. \ref{fig:schm}. Because $b_c$ might depend on the signs $s_1,~s_2$, the resulting impact parameters $b_n$ are also expected to contain $s_1$ and $s_2$.

With known impact parameters, the apparent angles of the series images is then given by Eq. (2.7) and (2.9) of Ref. \cite{Huang:2020trl} (after a typo corrected there)
\begin{align}
\theta_n(s_1,s_2)   =\arcsin\left[\frac{2s_2 b_nA(r_d)v-B(r_d)}{s_1 H(r_d)}\right] .\label{eq:thetasol}
\end{align}

The solved impact parameters $b_n$, after using the total travel time \eqref{eq:sedt}, also allow us to find the time delay between different relativistic images. For trajectories with the same rotation directions but different numbers of loops around the lens, $n_1$ and $n_2$, this is given by
\begin{align}
    \Delta^2 t_{n_1,n_2}=\Delta t(b_{n_2})-\Delta t(b_{n_1}).
\end{align}
Substituting Eq. \eqref{eq:sedt} and \eqref{eq:bnsol}, we can find that to the lowest order,
\begin{align}
    \Delta^2 t_{\uparrow n_1,\uparrow n_2}\approx&\frac{C_{t,0}}{C_{\phi,0}} \lcb \lsb (2n+1)s_1+1\rsb \pi-\Delta\phi_{sd}-D_{\phi,0}\rcb\Big|_{n=n_1}^{n=n_2}\nn\\
    =& 2(n_2-n_1)s_1\pi g_0(0)~~~~~~(n_1,n_2\in \mathbb{Z}_>),
    \label{eq:tdsamesidegeneral}
\end{align}
where in the last step Eqs. \eqref{eq:calpha0} and \eqref{eq:ytinyphi} are used.
Note that the constant term $D_{t,0}$ in the total travel time, although could be large if $r_{s}$ or $r_d$ is large, does not contribute to this time delay. If the signal is null, then $g_0(0)$ reduces to $s_2b_c$ and this formula recovers the corresponding formula in Ref. \cite{Bozza:2003cp}. In Kerr spacetime for null rays, this agrees with Refs. \cite{Kuang:2022xjp,Hsieh:2021rru,Wang:2016paq,Chen:2016hil,Islam:2021dyk,Islam:2021ful}.
Therefore, our result in Eq. \eqref{eq:tdsamesidegeneral} extends their results to signals with arbitrary velocity.  If the spacetime is static and spherically symmetric, then the factor $g_0(0)$ in Eq. \eqref{eq:tdsamesidegeneral} can be shown through its definition \eqref{eq:gn} to reduce to the product of a local time duration factor and the redshift factor from the critical radius to a far away observer, as revealed in Ref. \cite{Liu:2021ckg}.

For two signals rotating along different directions of the lens, assuming the one with direction $(s_1,s_2)$ rotates $n_1$ times and the one with direction $(-s_1,-s_2)$ rotates $n_2$ times, then their time delay is \begin{align}
    &\Delta^2 t_{\uparrow n_1,\downarrow n_2}\nn\\
    =&g_0(-s_1,-s_2)\left\{\left[ -(2n_2+1)s_1+1\right] \pi-\Delta\phi_{sd}\right.\nn\\
    &\left.-D_{\phi,0}(-s_1,-s_2)\right\}+D_{t,0}(-s_1,-s_2)\nn\\
    &-g_0(s_1,s_2)\left\{\lsb (2n_1+1)s_1+1\rsb \pi-\Delta\phi_{sd}\right.\nn\\
    &\left.-D_{\phi,0}(s_1,s_2)\right\}-D_{t,0}(s_1,s_2)~~(n_1,n_2\in \mathbb{Z}_>). \label{eq:tdoppsidegeneral}
\end{align}

\section{Applications to particular spacetimes\label{sec:apptoss}}

\subsection{Kerr spacetime}

We first consider the easiest SAS spacetime, Kerr spacetime, described by
\begin{align}
    \mathrm{d}s^2=&-\left(1-\frac{2Mr}{\rho^2}\right)\dd t^2+\left[\Delta_\mathrm{K}+\frac{2Mr(r^2+\hat{a}^2)}{\rho^2}\right]\sin^2{\theta}\dd\phi^2\nn\\
    &-\frac{4\hat{a}Mr\sin^2\theta}{\rho^2}\dd t \dd\phi+\frac{\rho^2}{\Delta_\mathrm{K}}\dd r^2+\rho^2\dd\theta^{2},
\end{align}
where $M$ is the spacetime mass and $\hat{a}=J/M$ is the spin angular momentum per unit mass of the Kerr BH, and
\begin{align}
    \rho^2=r^2+\hat{a}^2\cos^2\theta,~
    \Delta_\mathrm{K}(r)=r^2-2Mr+\hat{a}^2.
\end{align}
In the equatorial plane ($\theta=\pi/2$), this reduces to
\begin{align}
    \dd s^2=&-\left(1-\frac{2M}{r}\right)\dd t^2-\frac{4\hat{a}M}{r}\dd t\dd\phi\nn\\
    &+\left(r^2+\hat{a}^2+\frac{2\hat{a}^2M}{r}\right)\dd\phi^2 +\frac{r^2}{r^2-2Mr+\hat{a}^2}\dd r^2,
    \label{eq:kerr}
\end{align}
from which we can easily read off the metric functions $A(r),~B(r),~C(r)$ and $D(r)$ in this case.

\begin{figure}[htp!]
\centering
\includegraphics[width=0.45\textwidth]{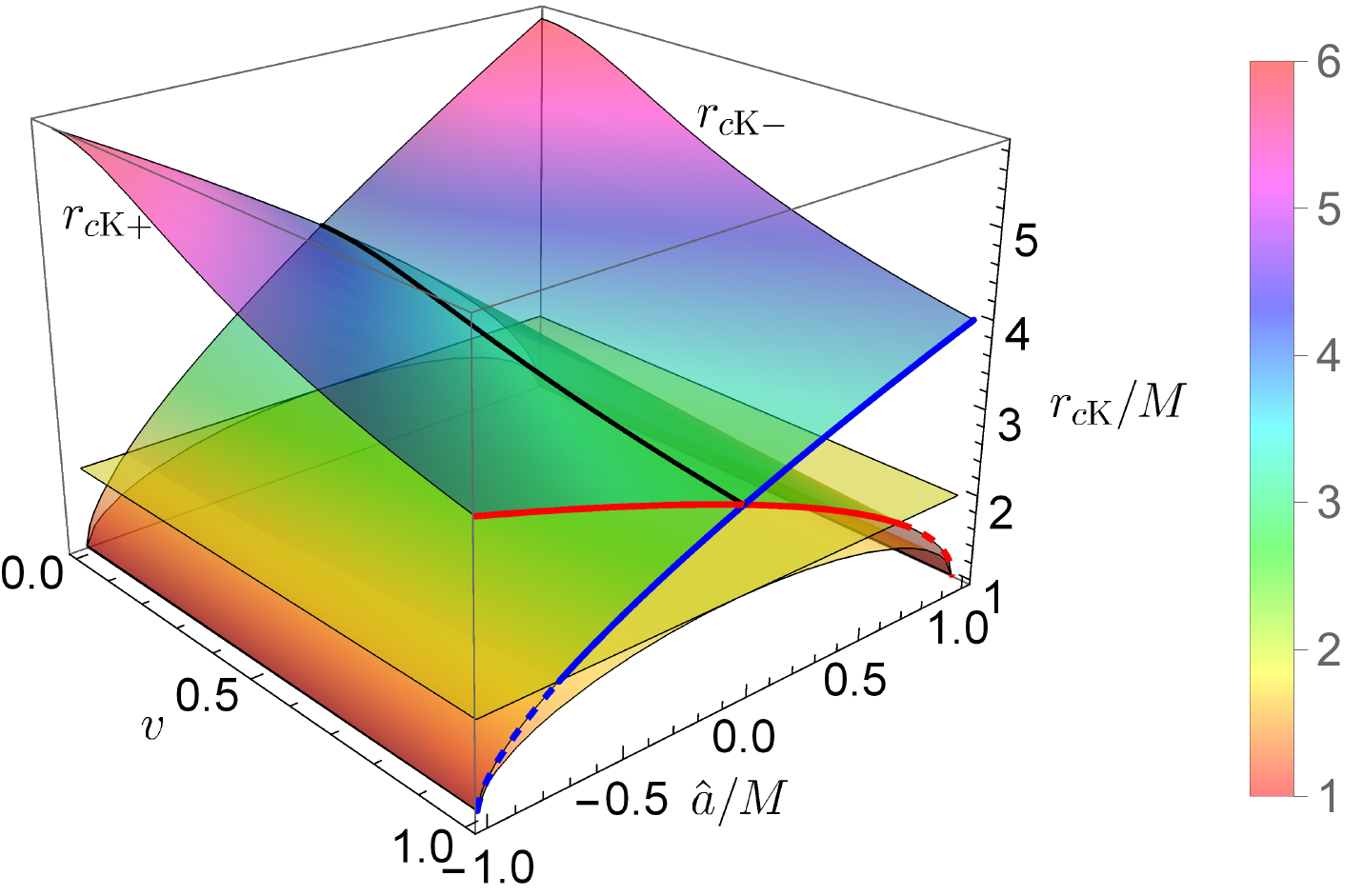}\\
(a)\\
\includegraphics[width=0.45\textwidth]{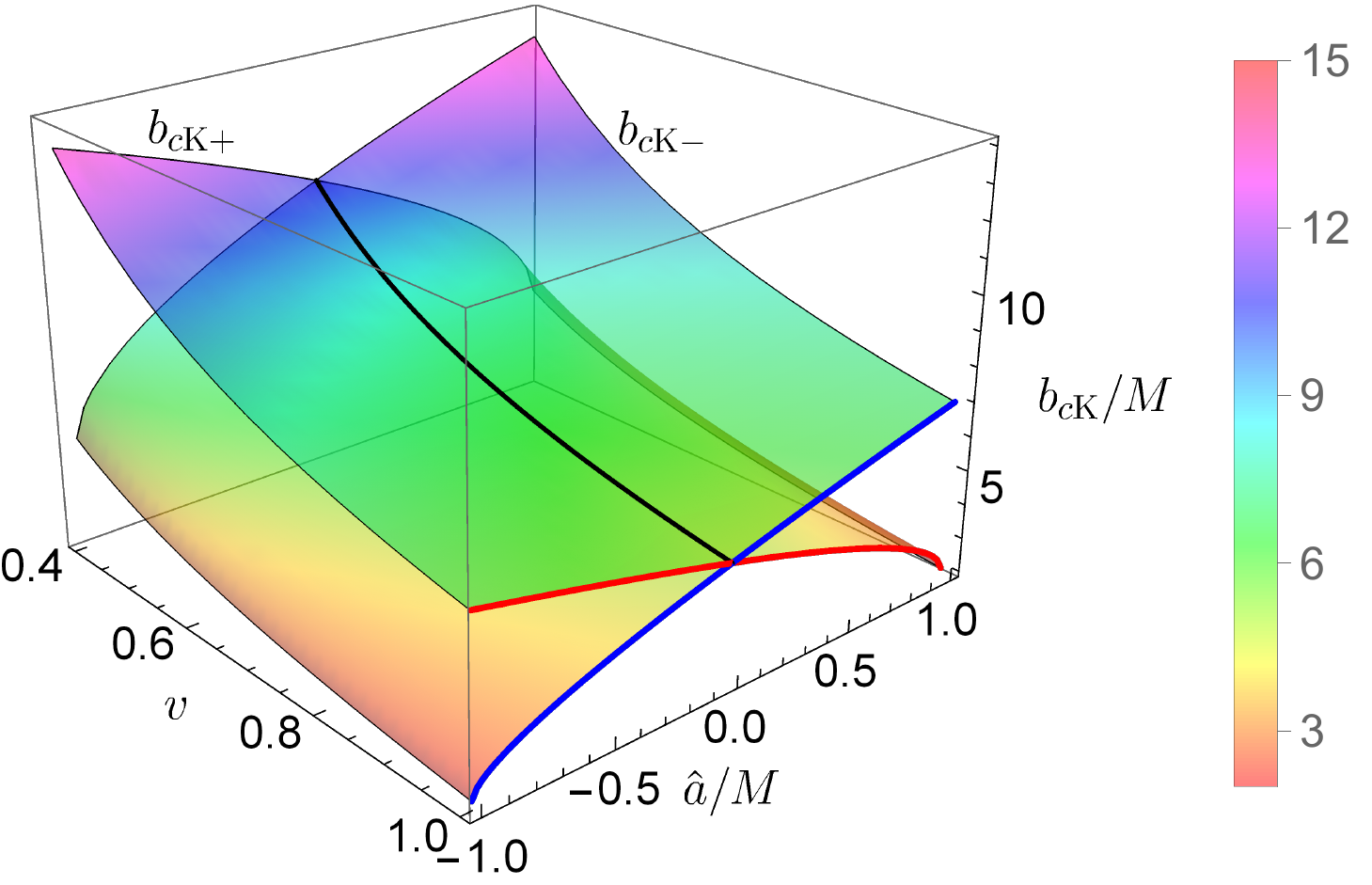}\\
(b)
\caption{\label{fig:rcbckerr} The critical $r_{c\mathrm{K}\pm}$ (a) and $b_{c\mathrm{K}\pm}$ (b) for the Kerr BH as functions of $\hat{a}$ and signal velocity $v$. In (a), we also plot the radii $r_{h\mathrm{K}}$ of the BH horizon and $r_{e\mathrm{K}}$ of the ergosurface. }
\end{figure}

Then in order to compute $\Delta\phi$ and $\Delta t$, we shall calculate $r_{c}$ and $b_{c}$ first. Using Eqs. \eqref{eq:dr=0}, \eqref{eq:definerc} and \eqref{eq:kerr}, we find that $r_c$ in Kerr spacetime is the solution of the following six-order polynomial
\begin{align}
    &v^4 r_c^6+2M \left(v^2-4v^4\right)r_c^5+M^2 \left(1-16v^2+24v^4\right)r_c^4\nonumber\\
    &+2M\left[4M^2(-1+5v^2-4v^4)-\hat{a}^2(v^2+v^4)\right]r_c^3\nonumber\\
    &+2M^2 \left[(8M^2+4\hat{a}^2)(1-v^2)^2-5\hat{a}^2(1-v^2)\right]r_c^2 \nonumber\\
    &-8\hat{a}^2M^3(1-v^2)^2r_c+\hat{a}^4M^2(1-v^2)^2=0.\label{eq:krceq}
\end{align}
Unfortunately, this equation does not permit explicit and analytical  solutions except for some special cases, such as $\kappa=0$ where the solutions are given in Eq. \eqref{eq:rckerrk0}. However, it is easy to solve numerically and we can show that it always allows two physical solutions, $r_{c\mathrm{K}\pm}$, corresponding to the two approaching directions of the signals with respect to the positive $z$ direction. These two roots are plotted as functions of $\hat{a}$ and $v$ in Fig. \ref{fig:rcbckerr} (a), as two crossed surfaces ($r_{c\mathrm{K}+}$ with red boundary at $v=1$ and $r_{c\mathrm{K}-}$ with blue boundary). We also illustrate the ergosurface radius $r_{e\mathrm{K}}$ and the BH radius $r_{h\mathrm{K}}$ in this figure. We observe that for all $(|\hat{a}|<M,v)$, the $r_{c\mathrm{K}\pm}>r_{h\mathrm{K}}$ while $r_{e\mathrm{K}}$
could be larger than $r_{c\mathrm{K}\pm}$ when $|\hat{a}|$ is large.
Moreover, when the signal is retrograde (the part of the surfaces above their black intersection line), the critical radius is always larger than that of the prograde one (the part of the surfaces below their intersection).

Once $r_{c\mathrm{K}\pm}$ are known, substituting them into Eq. \eqref{eq:1b_rc} for the Kerr case, we are able to work out the critical impact parameters $b_{c\mathrm{K}\pm}$. As stated in Sec.
\ref{sec:pm}, for $r_{c\mathrm{K}+}$, we found that  in Eq. \eqref{eq:1b_rc} the $s_1$ and $s_2$ have to be positive while for $r_{c\mathrm{K}-}$ they are both negative. Hence, we obtain the following using Eq. \eqref{eq:1b_rc}
\begin{align}\label{eq:kerrbc}
    &b_{c\mathrm{K}\pm}=\frac{\mp 2\hat{a} M+r_{c\mathrm{K}\pm} H_\mathrm{K}(r_{c\mathrm{K}\pm})/2}{ (r_{c\mathrm{K}\pm}-2M)v},
\end{align}
where $H_\mathrm{K}(r_{c\mathrm{K}\pm})$ is the $H(r)$ in Eq. \eqref{eq:hrdef} with the Kerr metric substituted, i.e.,
\begin{align}
    H_\mathrm{K}(r)=2\sqrt{\Delta_{\mathrm{K}}(r)\left[v^2+(1-v^2)\frac{2M}{r}\right]}.
\end{align}
Fig. \ref{fig:rcbckerr} (b) displays these two critical impact parameters, which have a qualitatively similar dependence on $\hat{a}$ and $v$ as the critical radii.

Using the metric functions in Eq. \eqref{eq:kerr}, we can work out their expansions near $r=r_{c\mathrm{K}\pm}$. To the order $\calco(r-r_{c\mathrm{K}\pm})^1$  they are
\begin{subequations}\label{eq:metric_rc}
\begin{align}
    A(r)=&\frac{r_c-2 M}{r_c}+\frac{2M(r-r_c)}{r_c^2}+\cdots,\\
    B(r)=&-\frac{4\hat{a} M}{r_c}+\frac{4\hat{a} M(r-r_c)}{r_c^2}+\cdots,\\
    C(r)=&\frac{2\hat{a}^2 M+\hat{a}^2 r_c+r_c^3}{r_c}\nn\\
    &+\frac{2(r_c^3-\hat{a}^2 M )(r-r_c)}{r_c^2}+\cdots,\\
    D(r)=&\frac{r_c^2}{r_c^2-2Mr_c+\hat{a}^2}\nn\\
    &+\frac{2r_c(\hat{a}^2-M r_c)(r-r_c)}{(r_c^2-2Mr_c+\hat{a}^2)^2}+\cdots,
\end{align}
\end{subequations}
where and henceforth in this section, $r_c$ is either $r_{c\mathrm{K}+}$ or $r_{c\mathrm{K}-}$ unless otherwise stated, and similarly $b_c$ will stand for $b_{c\mathrm{K}\pm}$. From the above expansion, it is easy to read off the coefficients $a_i,b_i,c_i,d_i~(i=0,1)$. Using Eqs. \eqref{eq:pp} and \eqref{eq:kerr}, we obtain the expression for $p(x)$ in Kerr spacetime
\begin{align}
    &p_{\mathrm{K}}(x)=\frac{1}{b_c}+\frac{s_2(1-2Mx)v}{2\hat{a}Mx-s_1 H_\mathrm{K}(1/x)/2}.
\end{align}
The inverse function $q_{\mathrm{K}}(x)$ of $p_{\mathrm{K}}(x)$ can be easily obtained by expanding the above $p_{\mathrm{K}}(x)$ and substituting the expansion coefficients into Eq. \eqref{eq:qxseries}.
Then, further substituting the coefficients in Eq. \eqref{eq:metric_rc}, as well as those of $p_{\mathrm{K}}(x)$ and $q_{\mathrm{K}}(x)$ into Eqs. \eqref{eq:yphinfirstfew}-\eqref{eq:ytningn} and then into Eq. \eqref{eq:cdfirstfew}, we are able to obtain the deflection angle and travel time in the Kerr BH spacetime to the leading order as
\begin{align}
    \Delta\phi_{\mathrm{K}\pm}&=C_{\mathrm{K}\phi,0}\ln{a_\pm}+D_{\mathrm{K}\phi,0}+\mathcal{O}(a_\pm)^1,\label{eq:kdefres}\\
    \Delta t_{\mathrm{K}\pm}&=C_{\mathrm{K}t,0}\ln{a_\pm}+D_{\mathrm{K}t,0}+\mathcal{O}(a_\pm)^1,\label{eq:kdtimeres}
\end{align}
where the small variable $a_\pm=1-b_{c\mathrm{K}\pm}/b$ and the coefficients  $C_{\mathrm{K}\phi,0},~D_{\mathrm{K}\phi,0},~C_{\mathrm{K} t,0},~D_{\mathrm{K} t,0}$
are formally still given by Eqs. \eqref{eq:calpha0} and \eqref{eq:cdfirstfew}
but with the following coefficients specialized in the Kerr spacetime
\begin{subequations}
    \begin{align}
    y_{\phi,-1}=&-\frac{s r_c^3 \zeta_h  q_1}{4\Delta_\mathrm{K}(r_c) b_c}\sqrt{\frac{ H_\mathrm{K}(r_c)}{ v}},\\
    y_{\phi,0}=&\frac{y_{\phi,-1}}{\zeta_h}+\frac{ s r_c^5 \zeta_h  q_1^2}{2 \Delta_\mathrm{K}(r_c) b_c  }\sqrt{\frac{ H_\mathrm{K}(r_c)}{ b_cv}}\nn\\
    &\times\left[\frac{1}{r_c}-\frac{q_2}{r_c^2 q_1^2}-\frac{\hat{a} M}{s r_c H_\mathrm{K}(r_c)(r_c-2M)}\right.\nn\\
    &\left.-\frac{(r_c-M)(2\hat{a}^2-\Delta_\mathrm{K}(r_c))}{2(\hat{a}^2-\Delta_\mathrm{K}(r_c))\Delta_\mathrm{K}(r_c)}\right],\\
   g_0=&2\frac{r_c\Delta_\mathrm{K}(r_c)-s\hat{a}MH_\mathrm{K}(r_c)}{sH_\mathrm{K}(r_c)(r_c-2M)},\\
   g_1=&-\frac{8q_1}{\sqrt{b_c}H_\mathrm{K}^2(r_c)}\left[\hat{a}^3M+3\hat{a}Mr_c^2\right.\nn\\
   &\left.+s (-2\hat{a}^2M-3Mr_c^2+r_c^3)b_c v+\hat{a}M b_c^2v^2\right],\\
    \zeta_h=&\sqrt{1+\frac{s b_c v(\sqrt{M^2-\hat{a}^2}-M)}{2\hat{a}M}}.
\end{align}
\end{subequations}
We remind the readers that the $r_c$ and $b_c$ in above equations are $r_{c\mathrm{K}\pm}$ and $b_{c\mathrm{K}\pm}$ respectively, and for the $+$ sign we have $s_1=s_2=s=+1$ and for the $-$ sign $s_1=s_2=s=-1$.

\begin{figure}[htp!]
\centering
\includegraphics[width=0.45\textwidth]{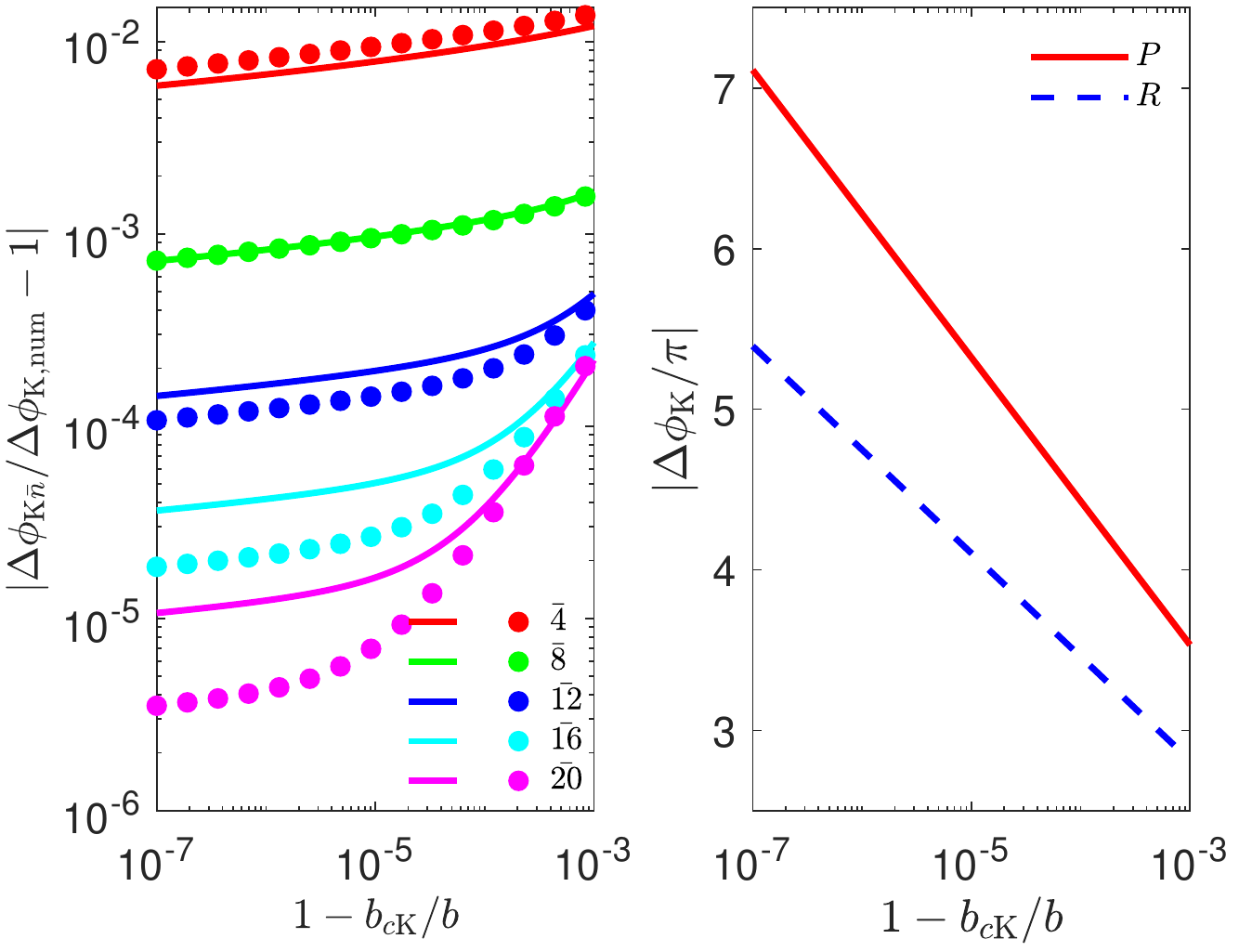}
\caption{\label{fig:kerrdphidt} (left) The difference between perturbative $\Delta\phi_{\mathrm{K}\bar{n}}$ and numerical $\Delta\phi_{\mathrm{K,num}}$ with solid lines for prograde motions and dots for retrograde motions, and (right) the $\Delta\phi_{\mathrm{K} \bar{20}}$ itself as a function of $1-b/b_c$ for prograde (P) and retrograde (R) motions. We set $\hat{a}=0.4M$ in this figure. }
\end{figure}

For the deflection angle \eqref{eq:kdefres}, we attempt to compare them with the values obtained by numerically integrating Eq. \eqref{eq:dphidef} for some chosen values of $\hat{a}$ and other parameters. However, through this process, we found that the expansions \eqref{eq:yphindef} for small $\xi=1-b_c/b(r)$ do not have a convergence radius in the parameter space covering all $|\hat{a}|\leq M$. Instead, we found that if $r_{c\mathrm{K}+}\leq r_{e\mathrm{K}}$ the ergospurface radius for the prograde motion and a given $(v,~\hat{a})$, there will exist a singularity point in the unphysical region of $\xi$ that blocks this half-integer series from converging. For example, for prograde null signals, if $\hat{a}\geq \sqrt{2}M/2$, then as one can see from Fig. \ref{fig:kerrdphidt} the $r_{c+}$ will shrink into the ergosurface radius. For this reason, the expansion result \eqref{eq:kdefres} for  a prograde (or retrograde) signal will only converge for $\hat{a}\in [-M, \hat{a}_e)$ (or $\hat{a}\in ( -\hat{a}_e,~M]$) where $\hat{a}_e$ is the limit determined by the intersection of $r_{c\mathrm{K}+}$ (or $r_{c\mathrm{K}-}$) and $r_{e\mathrm{K}}$. Note that this does not mean that in the prograde case with $\hat{a}\geq \hat{a}_e$ there will be no SDL. We verified
numerically that the prograde motion trajectory seems locally normal around $r\approx r_{e\mathrm{K}}$. Therefore, we believe that this convergence radius in $\hat{a}$ is only due to our particular expansion method, but not a physical singularity. In the following numerical studies, we will then restrict ourselves to the spacetime spin described above.

Now we can confirm the validity of result \eqref{eq:kdefres} by comparing its values $\Delta\phi_{\mathrm{K}\bar{n}}$, where $\bar{n}$ is the truncation order in the evaluation of $D_{\mathrm{K}\phi,0}$, with the $\Delta\phi_{\mathrm{K,num}}$  obtained by numerically integrating Eq. \eqref{eq:dphidef} for the Kerr spacetime. The numerical value can be thought as the true value of the deflection angle as long as the integrations are evaluated to high enough accuracy.
In Fig. \ref{fig:kerrdphidt} (left),  we plot the difference $\left|\Delta\phi_{\mathrm{K}\pm}/\Delta\phi_{\mathrm{K,num}}-1\right|$
as functions of $a=1-b_{c\mathrm{K},\pm}/b$ for fixed $\hat{a}=0.4M,~v=1$. For $M$ and $r_d$ in this and all following figures, we fix their values as those of the Sgr A$^*$, i.e.,  $M=4.297\times10^6 M_{\odot},~r_d=8.277$ [kpc] \cite{GRAVITY:2021xju} and assume $r_s=r_d$. It is seen that as the truncation order $\bar{n}$ increases, the perturbative result approaches the numerical value exponentially. Additionally, as $a$ decreases, the difference also decreases monotonically to extremely small values. This plot confirms the validity of Eq. \eqref{eq:kdefres}.
In Fig. \ref{fig:kerrdphidt} (right), the $\Delta \phi_{\mathrm{K}\pm}$  themselves with truncation order $\bar{n}=20$ are plotted. It is seen that even at the same $a$, the prograde trajectory has a larger deflection angle than the retrograde trajectory, mainly due to the fact that the former has a smaller critical radius, i.e., $r_{c\mathrm{K}+}<r_{c\mathrm{K}-}$. The dependence of $\Delta\phi_\mathrm{K}$ on $a$ in this figure is found to be consistent with Ref. \cite{Wang:2016paq}.

\begin{figure}[htp!]
\centering
\includegraphics[width=0.45\textwidth]{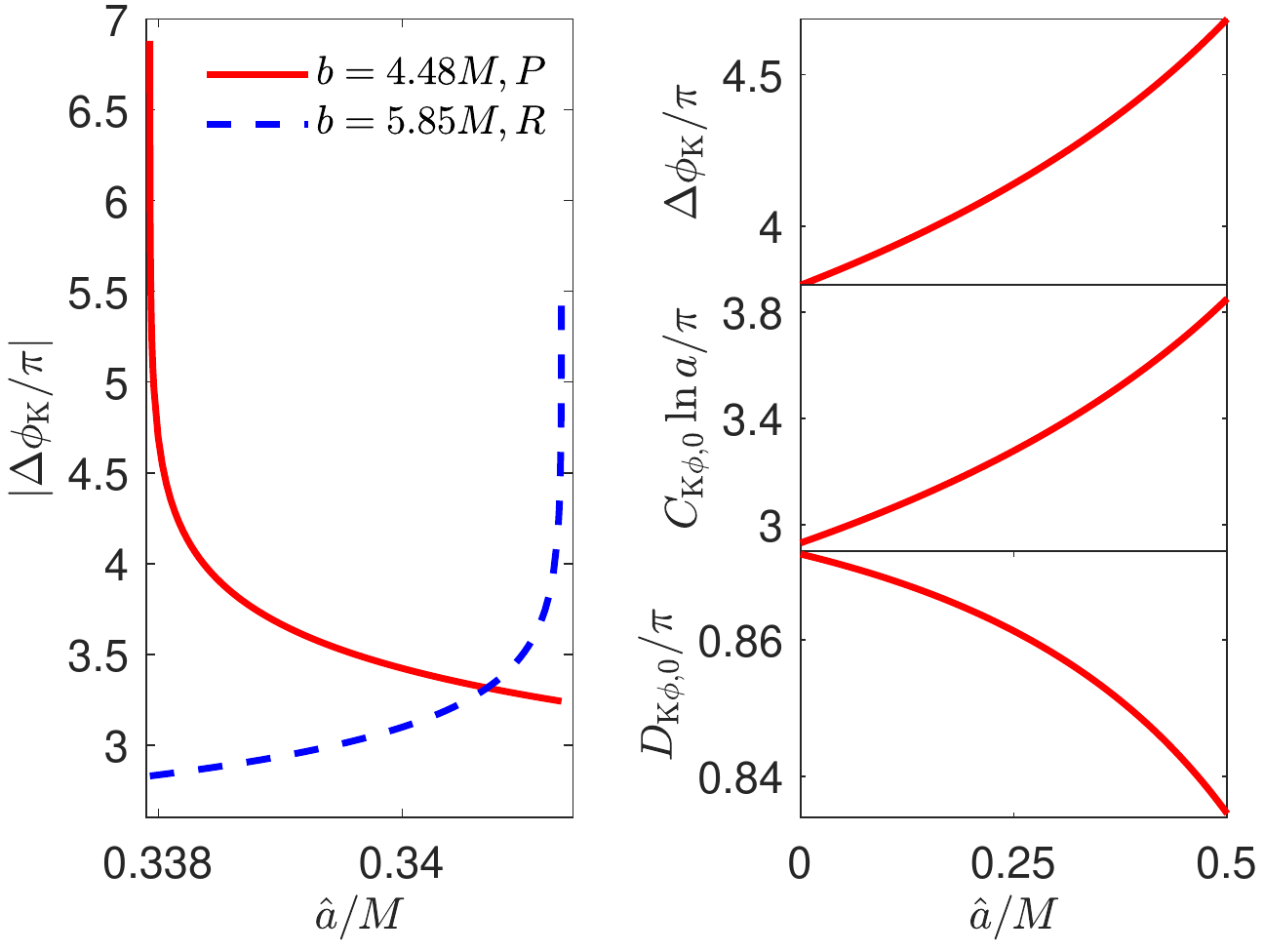}
\caption{\label{fig:kerrdphifora} (left) The $\Delta\phi_{\mathrm{K}}$ as a function of spin $\hat{a}$ with $b$ fixed at $4.48M$ and $5.85M$, slightly larger than $b_c$ from each direction; (right) Different terms' contribution to $\Delta\phi_{\mathrm{K}}$ as a function of $\hat{a}$ for prograde signal with fixed $1-b_c/b$ at $10^{-4}$. $P$ is for prograde and $R$ for retrograde motions. }
\end{figure}

In Fig. \ref{fig:kerrdphifora}, we investigate how the spacetime spin affects the deflection. In Fig. \ref{fig:kerrdphifora} (left), we fix the impact parameter and vary $\hat{a}$ within only a very small range such that the SDL of $a\ll 1$ is always satisfied.
It is seen that as $\hat{a}$ increases, the critical $b_{c\mathrm{K}}$ decreases (or increases) for prograde (or retrograde) motion which then makes the trajectory less (or more) critical, and therefore the deflection angle becomes smaller (or larger).
In Fig. \ref{fig:kerrdphifora} (right), we fix $a=1-b_{c\mathrm{K}\pm}/b$ while varying $\hat{a}$ over a larger range. It was clear from Eq. \eqref{eq:dphikbres} that the change in $\Delta\phi_{\mathrm{K}}$ is completely determined by the changes of coefficients $C_{\mathrm{K}\phi,0}$ and $D_{\mathrm{K}\phi,0}$. Then Fig. \ref{fig:kerrdphifora} (right) reveals that $|C_{\mathrm{K},0}|$ monotonically
increases as $\hat{a}$ increases, whereas $|D_{\mathrm{K},0}|$ monotonically decreases. In the limit of $a\to 0$, then $\Delta\phi_{\mathrm{K}}$ depends dominantly on  $|C_{\mathrm{K},0}|$, and consequently also increases as $\hat{a}$ increases. After setting $v=1$, we can also compare the dependence of result \eqref{eq:kdefres} on $\hat{a}$ with other works whose metric can be reduced to the Kerr metric. A quantitative comparison reveals that our result for light rays is consistent with Refs. \cite{Wei:2011nj,Chen:2016hil,Hsieh:2021scb,Kuang:2022xjp,Wang:2016paq}.

\begin{figure}[htp!]
\centering
\includegraphics[width=0.45\textwidth]{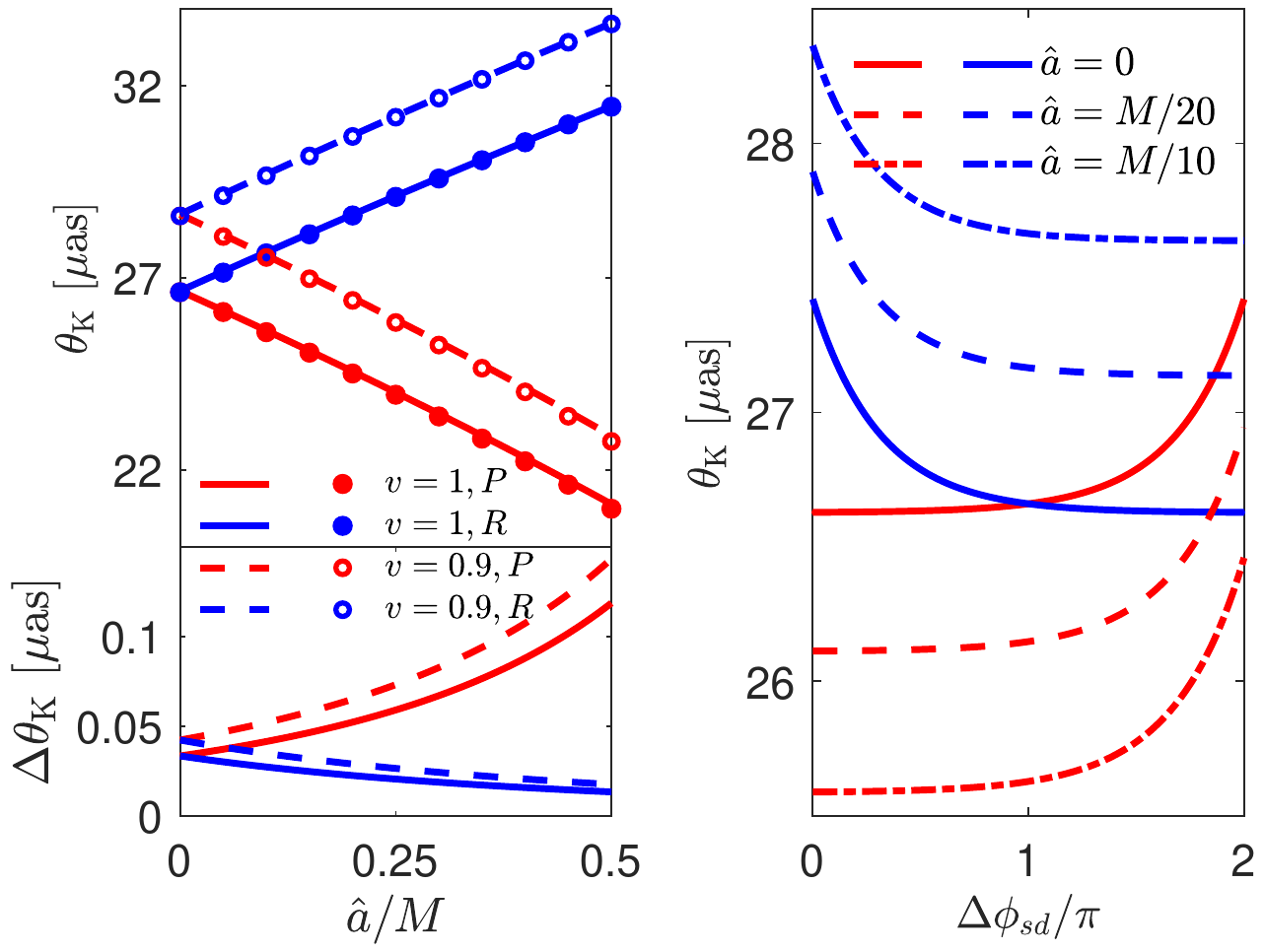}
\caption{\label{fig:kerrappang} The $\theta_{\mathrm{K}}$'s dependence on $\hat{a},~v$ (left) and $\Delta\phi_{sd},~\hat{a}$ (right) in Kerr spacetime. Top left: apparent angles for prograde (red curves and dots) and retrograde (blue curves and dots) motions for $n=1$ (solid curves) and $n=\infty$ (dots) with $v=1$ (solid curves and solid dots) and $v=0.9$ (dash curves and hollow dots); Bottom left: the separation of $\theta_1$ and $\theta_\infty$ (same legends as top left panel); Right: prograde (red curves) and retrograde (blue curves) with $n=1$ and $v=1$. }
\end{figure}

With the deflection angle in this spacetime verified, we can now study how the GLed images in the SDL are affected by various parameters.
In the left panel of Fig. \ref{fig:kerrappang}, we plot the apparent angles $\theta_1$ and $\theta_\infty$ as functions of the spacetime spin as well as the signal velocity, for both the prograde and retrograde motions, using Eq. \eqref{eq:thetasol}.
For the case of light rays, we compared our results with others' works that also studied the apparent angles in the Kerr spacetime but only for the nearly aligned  source-lens-detector configuration. Our results for these deflection angles agree quantitatively with Refs. \cite{Wei:2011nj, Chen:2012kn,Hsieh:2021scb,Chen:2016hil,Wang:2016paq,Islam:2021dyk,Islam:2021ful,Kuang:2022xjp} under these conditions.
From Fig. \ref{fig:kerrappang} (left), it is seen that the difference between $\theta_1$ and $\theta_\infty$ is very small ($\lesssim 0.14~[\mu\mathrm{as}]$ for signals with velocity larger than $0.9c$) for any signal rotation directions and spacetime spin. Therefore, it is expected that these differences will not be distinguishable in the near future. In contrast, the spacetime spin will affect the apparent angles dramatically, decreasing (or increasing) the apparent angle of the prograde (or retrograde) moving signal by about $5.5~[\mu\mathrm{as}]$ when $\hat{a}$ changes from zero to $0.5M$.
The asymptotic velocity also has a simple impact on the apparent angle. As the velocity decreases, the apparent angles of signals from both rotation directions will be increased. Qualitatively, the above features agree with the effect of $\hat{a}$ and $v$ on the critical impact parameters, as shown in Fig. \ref{fig:rcbckerr}. In the right panel of Fig. \ref{fig:kerrappang}, we show the dependence of the apparent angles on the source angular position for different spacetime spin $\hat{a}$. $\Delta\phi_{sd}=\pi$ corresponds to the location of the source aligned to the lens and observer. It is seen that the apparent angles of prograde (or retrograde) signals for a fixed spacetime spin will increase (or decrease) as $\Delta\phi_{sd}$ increases. As $\Delta\phi_{sd}$ increases from about $0$ to $2\pi$, the apparent angles changes by roughly 0.7$\sim$0.9~$[\mu\mathrm{as}]$.

\begin{figure}[htp!]
\centering
\includegraphics[width=0.45\textwidth]{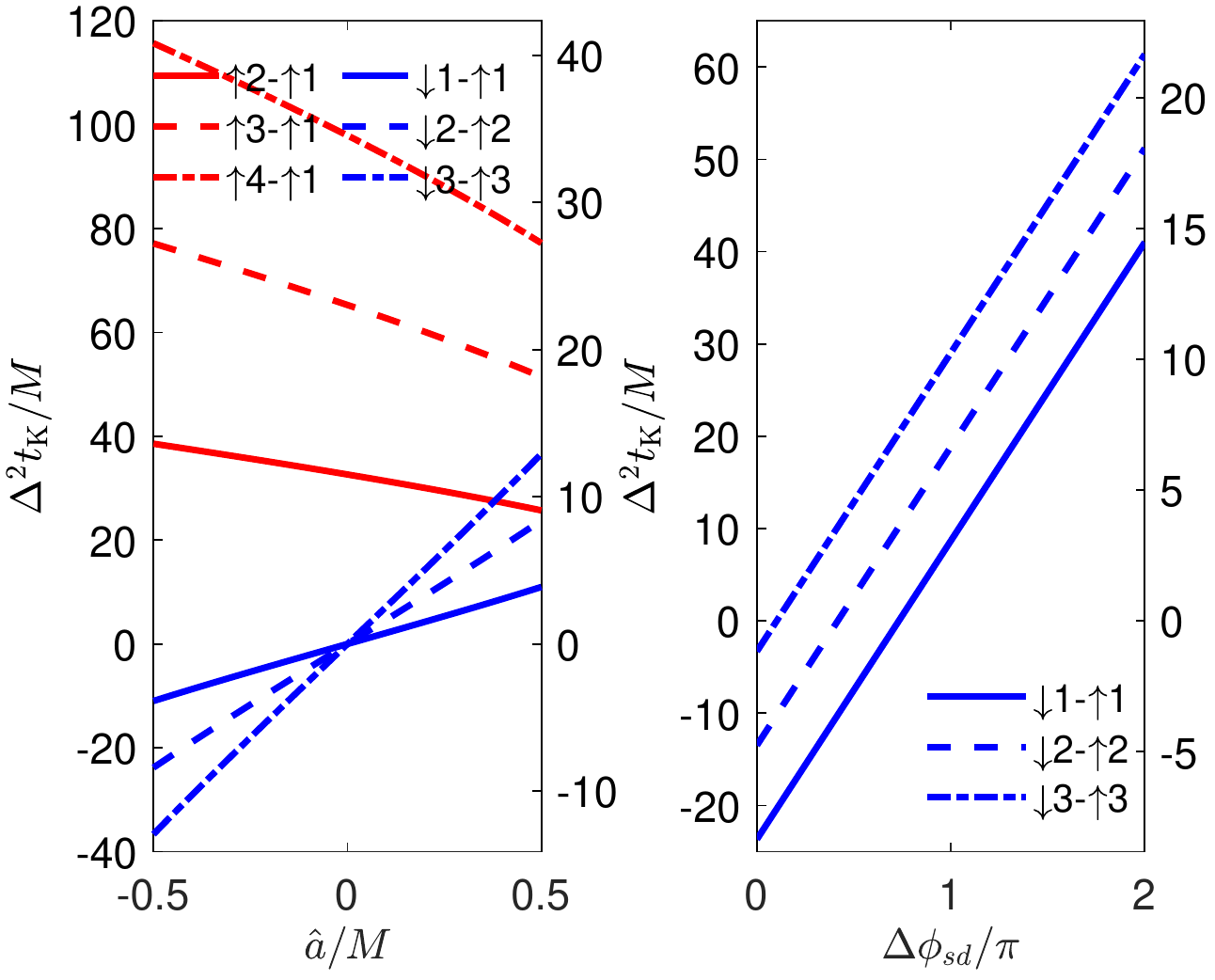}
\caption{\label{fig:kerrtdplot} Left: time delays as functions of the spacetime spin using Eq. \eqref{eq:tdsamesidegeneral} between signals anticlockwise rotating $n=2,3,4$ cycles and $n=1$ cycle (red curves), and  between signals looped the same $n=1$ or $n=2$ cycles but along different directions (retrograde minus prograde; blue curves) using Eq. \eqref{eq:tdoppsidegeneral}. Right: time delay's dependence on $\Delta\phi_{sd}$ for signals from different directions. We assumed $\Delta\phi_{sd}=\pi$ in the left panel and $\hat{a}=0.4M$ in the right panel and $v=1$ in all plots. }
\end{figure}

We can also study the time delay between images from  the same or opposite sides of the lens in Kerr spacetime using Eqs. \eqref{eq:tdsamesidegeneral} and \eqref{eq:tdoppsidegeneral}. The dependence of the time delays on $\hat{a}$ and $\Delta\phi_{sd}$ are plotted in Fig. \ref{fig:kerrtdplot}, whose ticks on the left and right $y$-axes are in the unit of $M$ and [min] respectively. In Fig. \ref{fig:kerrtdplot} (left, red curves), we plot the time delay $\Delta^2 t_{\uparrow n_1,\uparrow n_2}$ between signals looped from the same side of the lens for $n_2=2,~3,~4$ cycles and $n_1=1$ cycle. In this plot, we fix the motion direction of the signal to be anticlockwise so that the $\hat{a}<0$ part corresponds to the retrograde motion. It is seen that the time delay increases linearly as $(n_2-n_1)$ increases, as suggested by Eq. \eqref{eq:tdsamesidegeneral}. Moreover, as $|\hat{a}|$ increases from zero to about $0.5M$, the time delay between $n_2=2$ and $n_1=1$ for prograde (or retrograde) motion decreases (or increases) almost linearly from about 11.5 min to 9.1 min (or 13.6 min).
This is indeed a reflection of the dependence of $r_c$ on $\hat{a}$ as shown in Fig. \ref{fig:rcbckerr}.
In Fig. \ref{fig:kerrtdplot} (left, blue curves), the time delay $\Delta^2 t_{\uparrow n,\downarrow n}$ between signals winding the same $n$ cycles around the BH but from opposite sides of the lens are plotted according to Eq. \eqref{eq:tdoppsidegeneral}. Note that when computing $D_{t,0}(\pm s_1,\pm s_2)$ in this equation, in order for the series to converge faster, we cut out a large portion of the trajectory from $r_s,~r_d$ to a much smaller radius. This will not affect the time delay very much because the main contribution to the time delay happens around the particle sphere but not at larger radius. From the plot it is seen for the source located at $\phi_s=\pi$, the two rotation directions will be reflectively symmetric when $\hat{a}=0$ and therefore the time delay is zero at this point. As $\hat{a}$ grows to $0.5M$, all time delays increase linearly to about 4 [min] for each loop. Again, this dependence on $\hat{a}$ can be qualitatively understood from $r_c$ in Fig. \ref{fig:rcbckerr}: as $\hat{a}$ deviates from zero, the $r_{c\mathrm{K}-}$ increases and $r_{c\mathrm{K}+}$ decreases so that their difference increases, causing the time delay to grow. In Fig. \ref{fig:kerrtdplot} (right), the dependence of the same time delay $\Delta^2 t_{\uparrow n,\downarrow n}$ on $\Delta\phi_{sd}$ is plotted.
As $\Delta\phi_{sd}$ changes from $0$ to $2\pi$, the prograde signal loops one cycle less and the retrograde signal loops one cycle more, and therefore $\Delta^2 t_{\uparrow n,\downarrow n}$ (recalling this is the retrograde minus prograde travel times) increase by about the combined time to travel one cycle clockwise and one cycle anticlockwise. From the left subplot (solid curves) we can see that this combined time is about 24 [min], which is exactly the change for each line in this subplot.

\subsection{Rotating Kalb-Ramond BH spacetime}

In this section, we apply our perturbative method to solve for the $\Delta\phi$ and the GL equation in the SDL in the rotating KR BH spacetime, described by the line element  \cite{Kumar:2020hgm}
\begin{align}
    \dd s^2=&-\left(\frac{\Delta_{\mathrm{KR}}-\hat{a}^2\sin^2{\theta}}{\rho^2}\right)\dd t^2+\frac{\rho^2}{\Delta_{\mathrm{KR}}}\dd r^2\nn\\
    &-2\hat{a}\sin^2{\theta}\left(1-\frac{\Delta_\mathrm{KR}-\hat{a}^2\sin^2{\theta}}{\rho^2}\right)\dd t\dd\phi+\rho^2\dd\theta^2\nn\\
    &+\sin^2{\theta}\left[\rho^2+\hat{a}^2\sin^2{\theta}\left(2-\frac{\Delta_\mathrm{KR}-\hat{a}^2\sin^2{\theta}}{\rho^2}\right)\right]\dd\phi^2, \label{eq:krmetric}
\end{align}
where
\begin{align}
    \rho^2=r^2+\hat{a}^2\cos^2{\theta},~\Delta_{\mathrm{KR}}=r^2-2Mr+\hat{a}^2+\frac{\Gamma}{r^{-2+2/\lambda}}.
\end{align}
Here $M$ is the mass of the BH and $\hat{a}=J/M$ is the angular momentum per unit mass.
Comparing to the Kerr spacetime, the rotating KR spacetime introduces two new parameters, $\Gamma$ and $\lambda$. Setting $\lambda=0$, Eqs. \eqref{eq:krmetric} as well as \eqref{eq:KR} reduce to the corresponding forms in the Kerr metric, while setting $\lambda=1$ results in the Kerr-Newman (KN) metric with $\Gamma$ playing the role of charge squared. Therefore, $\lambda$ measures the transition of the spacetime from Kerr to KN and $\Gamma$ stands for some effective charge of the spacetime.

The BH event horizon in this spacetime is determined by $\Delta_{\mathrm{KR}}=0$, which is not analytically solvable for general values of $\lambda$. However, it can be shown that at least for some regions in the parameter space, there exist two BH horizons. For example, when $\lambda=2/3$, $\Delta_{\mathrm{KR}}=0$ can be transformed into a solvable cubic equation of $r$, yielding two horizons at
\begin{align}
    r=r_{h\mathrm{KR}\pm}(\Gamma), ~~(r_{h\mathrm{KR}-}\leq r_{h\mathrm{KR}+}) \label{eq:cubsol}
    \end{align}
provided the following conditions are satisfied
\begin{align}
    f_-\leq \Gamma\leq f_+,~~|\hat{\hat{a}}|\equiv |\hat{a}/M|\leq 2\sqrt{3}/3, \label{eq:condg}
\end{align}
where
\begin{align}
f_\pm=\frac{2}{27} \left[8-9 \hat{\hat{a}}^2\pm\left(4-3\hat{\hat{a}}^2\right)^{3/2}\right].
\end{align}

In the equatorial plane ($\theta=\pi/2$), the line element becomes
\begin{align}
    \dd s^2=&-\left(1-\frac{2M}{r}+\frac{\Gamma}{r^{2/\lambda}}\right)\dd t^2+2\hat{a}\left(-\frac{2M}{r}+\frac{\Gamma}{r^{2/\lambda}}\right)\dd t\dd\phi\nn\\
    &+\left(r^2+\hat{a}^2+\frac{2\hat{a}^2M}{r}-\frac{\hat{a}^2\Gamma}{r^{2/\lambda}}\right)\dd\phi^2\nn\\
    &+\frac{r^2}{r^2-2Mr+\hat{a}^2+\Gamma r^{2-2/\lambda}}\dd r^2,
    \label{eq:KR}
\end{align}
from which we can read off the metric functions $A(r),~B(r),~C(r)$ and $D(r)$.
The radius of the ergosurface on the equatorial plane of this spacetime is determined by the equation $A(r)=0$. Similar to the solution to the BH horizon radii, the ergosurface radius for general $\lambda$ can only be solved numerically too. For the special value of $\lambda=2/3$, the outer ergosurface radius can be solved as
\begin{align}
    r_{e\mathrm{KR}}=\frac{2}{3}M\left\{1+2\cos\left[\frac{1}{3}\arccos\left(1-\frac{27\Gamma}{16M^3}\right)\right]\right\}.
\end{align}

In order to determine the deflection in the SDL in this spacetime, we first calculate its critical radius and impact parameter. Using Eqs. \eqref{eq:dr=0}, \eqref{eq:definerc} and \eqref{eq:KR}, we can obtain the equation that determines the critical radius in this spacetime. Similar to the Kerr case, this equation will usually only be solvable numerically except for very particular parameter choices. Therefore, we will not list the equation here but we do find two physical critical values $r_{c\mathrm{KR}\pm}$.
Using Eq. \eqref{eq:1b_rc}, the critical impact parameter $b_{c\mathrm{KR}\pm}$ is related to $r_{c\mathrm{KR}\pm}$ by the following relation (in the reminder of this subsection, we will use $r_c$ and $b_c$ to stand for $r_{c\mathrm{KR}\pm}$ and $b_{c\mathrm{KR}\pm}$ unless stated explicitly)
\begin{align}
    b_{c\mathrm{KR}\pm}=\frac{\mp2\hat{a}\left(2M-\Gamma r_c^{1-2/\lambda}\right)+r_cH_{\mathrm{KR}}(r_c)}{2\left(r_c-2M+\Gamma r_c^{1-2/\lambda}\right)v},
\end{align}
where $H_{\mathrm{KR}}(r)$ is the $H(r)$ in Eq. \eqref{eq:hrdef} with the rotating KR metric substituted, i.e.,
\begin{align}
    &H_{\mathrm{KR}}(r)=2\sqrt{\Delta_{\mathrm{KR}}(r)\left[v^2+(1-v^2)\lb \frac{2M}{r}-\frac{\Gamma}{ r^{2/\lambda}}\rb\right]}.
\end{align}
If $\lambda\to 0$, this equation reduces to the Kerr case Eq. \eqref{eq:kerrbc} as expected.

Having solved for $r_c$, we can expand the metric functions around it. To the $\mathcal{O}(r-r_{c})^1$ order, they are
\begin{subequations}
\label{eq:KRmetricexpansion}
\begin{align}
        A(r)=&\left(1-\frac{2M}{r_c}+\frac{\Gamma}{r_c^{2/\lambda}}\right)\nn\\
        &+2\left(\frac{M}{r_c^2}-\frac{\Gamma}{\lambda r_c^{1+2/\lambda}}\right)(r-r_c)+\cdots,\\
        B(r)=&2\hat{a}\left(-\frac{2M}{r_c}+\frac{\Gamma}{r_c^{2/\lambda}}\right)\nn\\
        &+4\hat{a}\left(\frac{M}{r_c^2}-\frac{\Gamma}{\lambda r_c^{1+2/\lambda}}\right)(r-r_c)+\cdots,\\
C(r)=&\left(r_c^2+\hat{a}^2+\frac{2\hat{a}^2M}{r_c}-\frac{\hat{a}^2\Gamma}{r_c^{2/\lambda}}\right)\nn\\
        &+2\left(-\frac{\hat{a}^2M}{r_c^2}+r_c+\frac{\hat{a}^2\Gamma}{\lambda r_c^{1+2/\lambda}}\right)(r-r_c)+\cdots,\\    D(r)=&\frac{r_c^2}{r_c^2-2Mr_c+\hat{a}^2+\Gamma r_c^{2-2/\lambda}}\nn\\
&+\frac{2r_c[\hat{a}^2+r_c(-M+\Gamma r_c^{1-2/\lambda}/\lambda)]}{[\hat{a}^2+r_c(r_c-2M+\Gamma r_c^{1-2/\lambda})]^2}(r-r_c)+\cdots.
    \end{align}
\end{subequations}
From this equation, we can read off the coefficients $a_i,b_i,c_i,d_i~(i=0,1)$. Again, higher order coefficients are also easy to obtain but too lengthy to show here. Using Eqs. \eqref{eq:pp} and \eqref{eq:kerr}, we can then obtain the function $p(x)$ in the rotating KR spacetime
\begin{align}
    p_{\mathrm{KR}}(x)=\frac{1}{b_c}+\frac{2s_2\left(1-2Mx+\Gamma x^{2/\lambda}\right)v}{2\hat{a}\left(2Mx-\Gamma x^{2/\lambda}\right)-s_1H_{\mathrm{KR}}(1/x)}.
\end{align}

We can expand $p_{\mathrm{KR}}(x)$ around $1/r_c$ to obtain the coefficients in Eq. \eqref{eq:pxseries}.  Then further substituting these coefficients, as well as those in Eqs. \eqref{eq:KRmetricexpansion} into Eqs. \eqref{eq:yphinfirstfew} and then into Eq. \eqref{eq:cdfirstfew}, we can obtain the deflection angle and travel time for the rotating KR BH to the leading order as
\begin{align}
    \Delta\phi_\mathrm{KR}&=C_{\mathrm{KR}\phi,0}\ln{a}+D_{\mathrm{KR}\phi,0}+\mathcal{O}(a)^1,\label{eq:dphikbres}\\
    \Delta t_\mathrm{KR}&=C_{\mathrm{KR}t,0}\ln{a}+D_{\mathrm{KR}t,0}+\mathcal{O}(a)^1,\label{eq:dtkbres}
\end{align}
where again the coefficients $C_{\mathrm{KR}\phi,0},~D_{\mathrm{KR}\phi,0},~C_{\mathrm{KR} t,0}$ and $D_{\mathrm{KR} t,0}$
are given by Eqs. \eqref{eq:calpha0} and \eqref{eq:cdfirstfew}
with the following coefficients specialized in the KR spacetime
\begin{subequations}
    \begin{align}
        &y_{\phi,-1}(0)=-\frac{s r_c^3 \zeta_h  q_1}{4\Delta_\mathrm{KR}(r_c) b_c}\sqrt{\frac{ H_\mathrm{KR}(r_c)}{v}},\\
        &y_{\phi,0}(0)=\frac{y_{\phi,-1}(0)}{\zeta_h}+\frac{sr_c^5\zeta_hq_1^2}{2\Delta_{\mathrm{KR}}(r_c)b_c}\sqrt{\frac{H_{\mathrm{KR}}(r_c)}{b_cv}}\nn\\
        &\times\left\{\frac{1}{r_c}-\frac{q_2}{r_c^2 q_1^2}-\frac{\hat{a}\left(M-\Gamma r_c^{1-2/\lambda}/\lambda\right)}{sr_cH_{\mathrm{KR}}(r_c)\left(r_c-2M+\Gamma r_c^{1-2/\lambda}\right)}\right.\nn\\
        &\left.+\frac{\left[r_c-M+\Gamma r_c^{1-2/s}(1-1/\lambda)\right]\left[2\hat{a}^2-\Delta_{\mathrm{KR}}(r_c)\right]}{2\Delta_{\mathrm{KR}}(r_c)\left(r_c-2M+\Gamma r_c^{1-2/\lambda}\right)r_c}\right\},\\
        &g_0(0)=\frac{2r_c\Delta_{\mathrm{KR}}(r_c)-s\hat{a}H_{\mathrm{KR}}(r_c)\left(2M-\Gamma r_c^{1-2/\lambda}\right)}{sH_{\mathrm{KR}}(r_c)\left(r_c-2M+\Gamma r_c^{1-2/\lambda}\right)},\\
        &g_1(0)=-\frac{8q_1}{\sqrt{b_c}H_{\mathrm{KR}}^2(r_c)}\left\{\hat{a}^3M+3\hat{a}Mr_c^2+s \left(-2\hat{a}^2M\right.\right.\nn\\
        &\left.-3Mr_c^2+r_c^3\right)b_c v+\hat{a}M b_c^2v^2-\frac{\Gamma r_c^{1-2/\lambda}}{\lambda}\nn\\
        &\left.\times\left[\hat{a}^3+(\hat{a}-s b_c v)r_c^2(1+\lambda)-2s\hat{a}^2b_cv+\hat{a}b_c^2v^2\right]\right\}.
    \end{align}
\end{subequations}

\begin{figure}[htp!]
\centering
\includegraphics[width=0.45\textwidth]{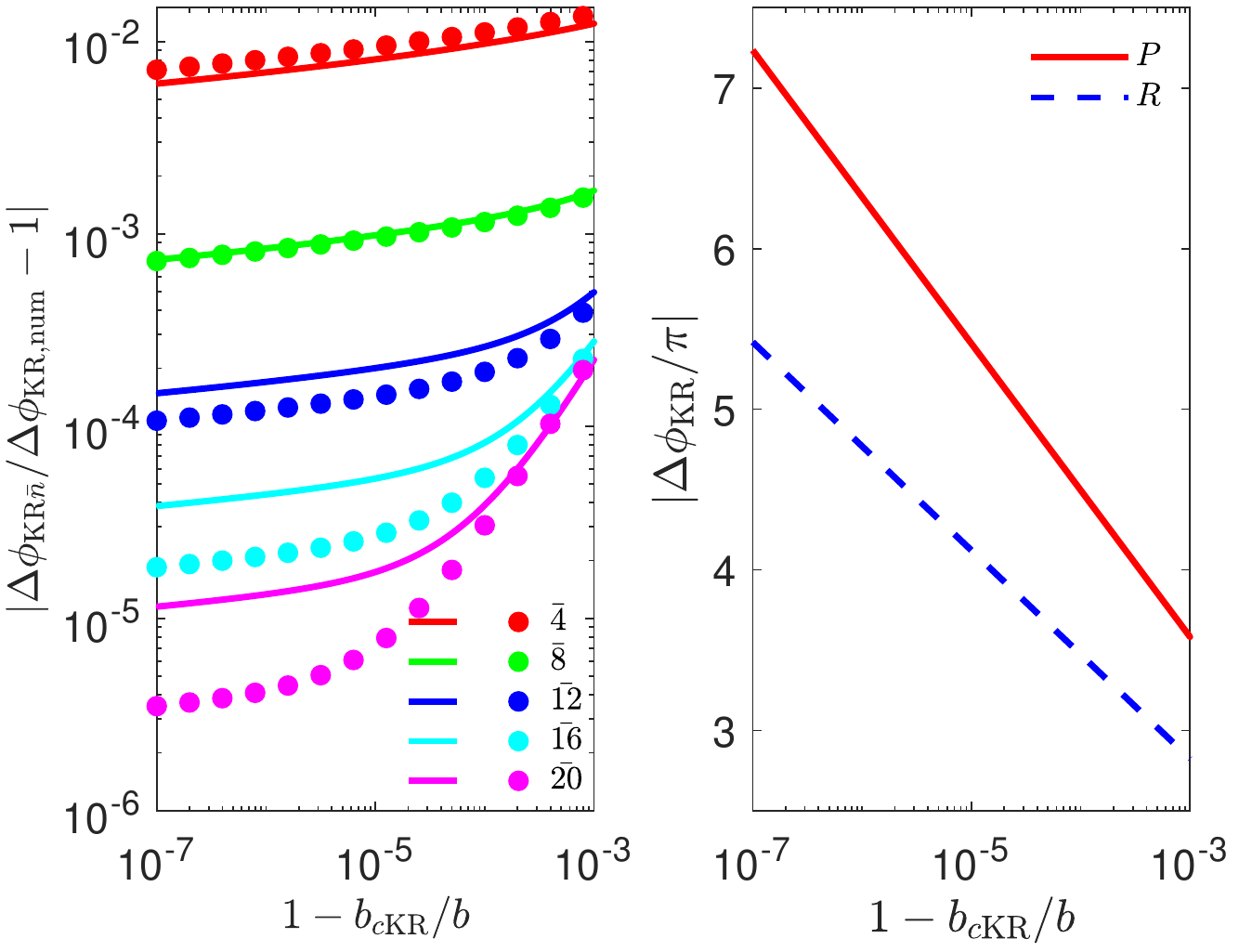}
\caption{\label{fig:krplot1} (left) The difference between perturbative $\Delta\phi_{\mathrm{KR}\bar{n}}$ and numerical $\Delta\phi_{\mathrm{KR,num}}$, and (right) the $\Delta\phi_{\mathrm{KR}\bar{20}}$ itself as a function of $1-b/b_c$ for two rotation directions. $P$ is for prograde and $R$ for retrograde motions. }
\end{figure}

To verify the results \eqref{eq:dphikbres}, in Fig. \ref{fig:krplot1} we plotted the truncated $\Delta\phi_{\mathrm{KR}}$ to order $\bar{n}$ and compared it with the numerical integration result. We chose $\hat{a}=0.4M,~\lambda=2/3$ and $\Gamma=0.1M^3$, and other parameters the same as in Fig. \ref{fig:kerrdphidt}. As shown in the figure, the perturbative results are in excellent agreement with the numerical integration results. Their difference decreases exponentially as the truncation order increases or as $b$ approaches $b_c$ from above.

\begin{figure}[htp!]
\centering
\includegraphics[width=0.45\textwidth]{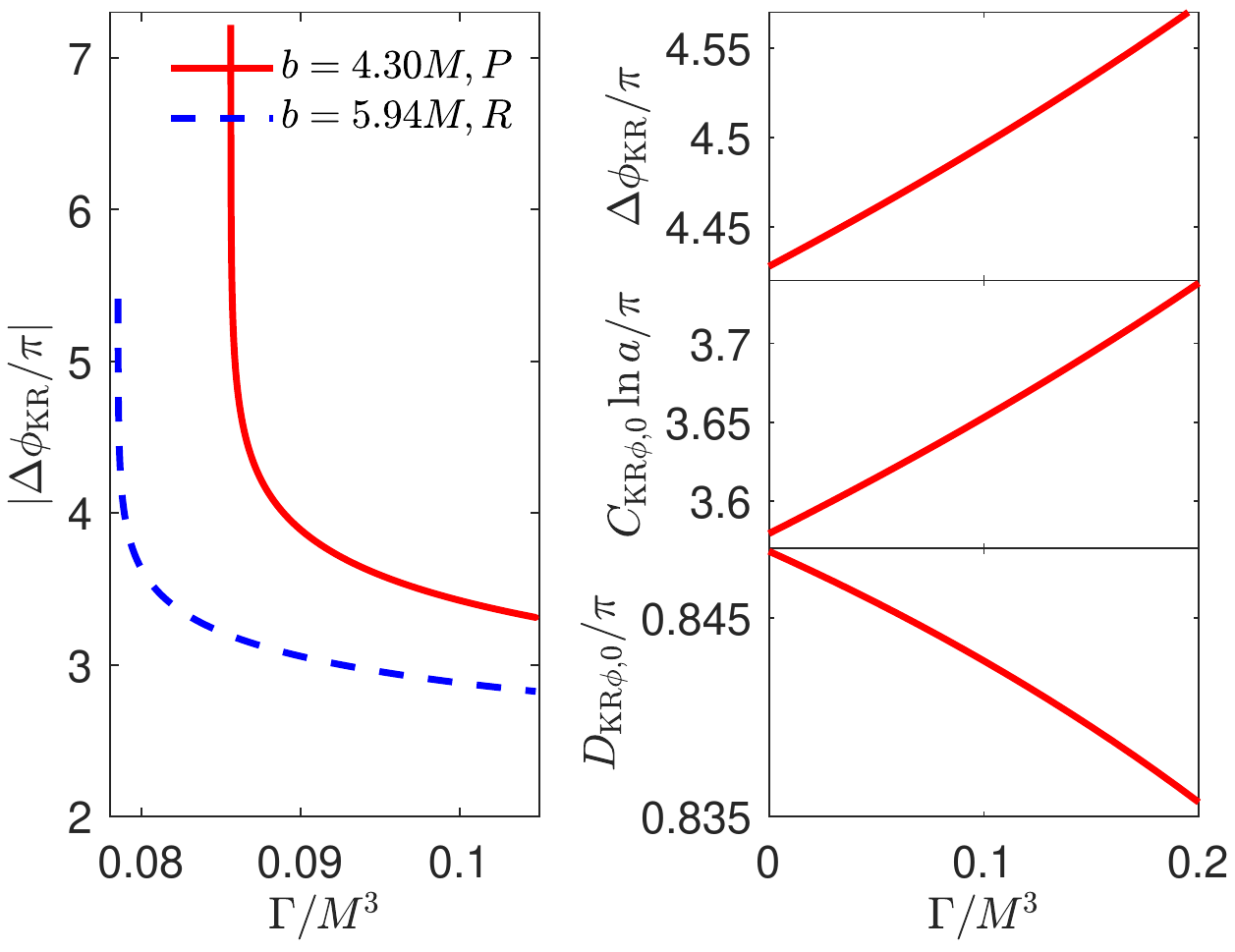}\\
(a)\\
\includegraphics[width=0.45\textwidth]{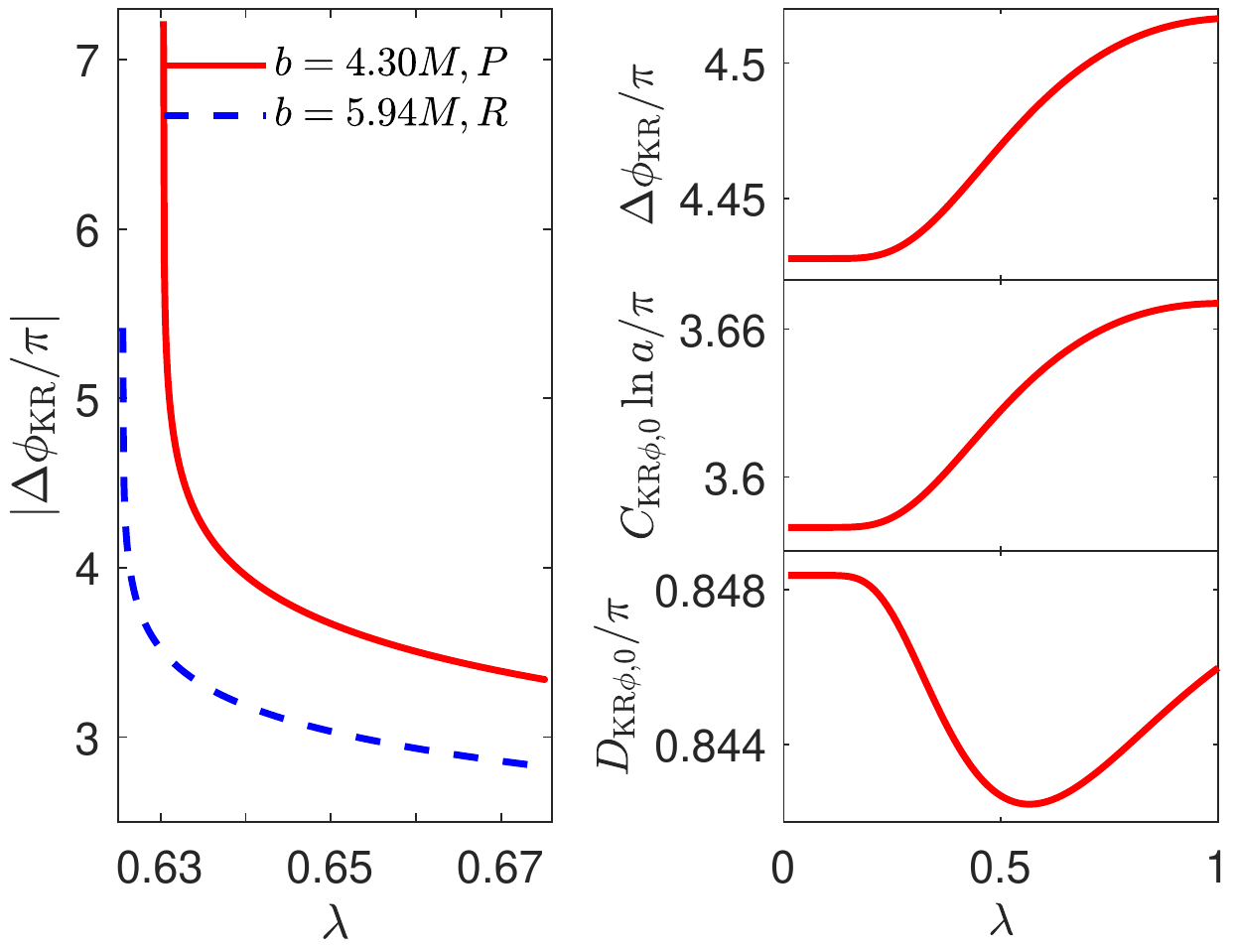}\\
(b)
\caption{\label{fig:krdphionother} The $\Delta\phi_{\mathrm{KR}}$ as a function of (a) $\Gamma$ and (b) $\lambda$ with (left) $b$ fixed to values slightly larger than $b_c$, and (right) $1-b_c/b$ fixed at $10^{-4}$. We fixed $\lambda=2/3$ in (a) and  $\Gamma=\Gamma_0 M^{2/\lambda}$ with $\Gamma_0=0.1$ in (b). Other parameters are the same as in Fig. \ref{fig:kerrdphidt}. $P$ is for prograde and $R$ for retrograde motions. }
\end{figure}

With the perturbative deflection angle's  correctness confirmed, in Fig. \ref{fig:krdphionother} we use them to study the effect of various spacetime parameters on the deflection. The spin $\hat{a}$'s effect is qualitatively similar to that in the Kerr spacetime as shown in Fig. \ref{fig:kerrdphifora}. Therefore, we will only focus on the effects of the parameters $\Gamma$ and $\lambda$ here. In Fig. \ref{fig:krdphionother} (a), we see that as the charge parameter $\Gamma$ increases, the deflection angle decreases rapidly, implying that the trajectories with fixed $b$  become less critical, i.e., both $b_{c\mathrm{KR}+}$ and $b_{c\mathrm{KR}-}$ decrease with increasing $\Gamma$. This is in contrast with the effect of the spin, whose increase always decreases the critical $b_c$ for one rotation direction while  increases that for the other direction, but in accord with the observation in Ref. \cite{Pang:2018jpm} that the increasing of the charge decreases the critical radius of the particle sphere. When $1-b_c/b$ is fixed, then we see that the increase of $\Gamma$ causes an increase in $C_{\mathrm{KR}\phi,0}\ln a$ and a weaker decrease in $D_{\mathrm{KR}\phi,0}$, whose combination still increases the deflection. This suggests that in the SDL with even larger $|\ln(a)|$, $\Gamma$ will always increase the deflection angle.
The effect of the deviation parameter $\lambda$ is shown in Fig. \ref{fig:krdphionother} (b). Since the deviation of $\lambda$ from 0 to 1 implies the transition of the spacetime from Kerr to KN, we see that this is also the process in which the charge of the KN spacetime becomes fully enabled. Therefore we can expect that increasing $\lambda$ should also effectively decrease the deflection angle, which agrees with the trend observed in this subplot.

\begin{figure}[htp!]
\centering
\includegraphics[width=0.45\textwidth]{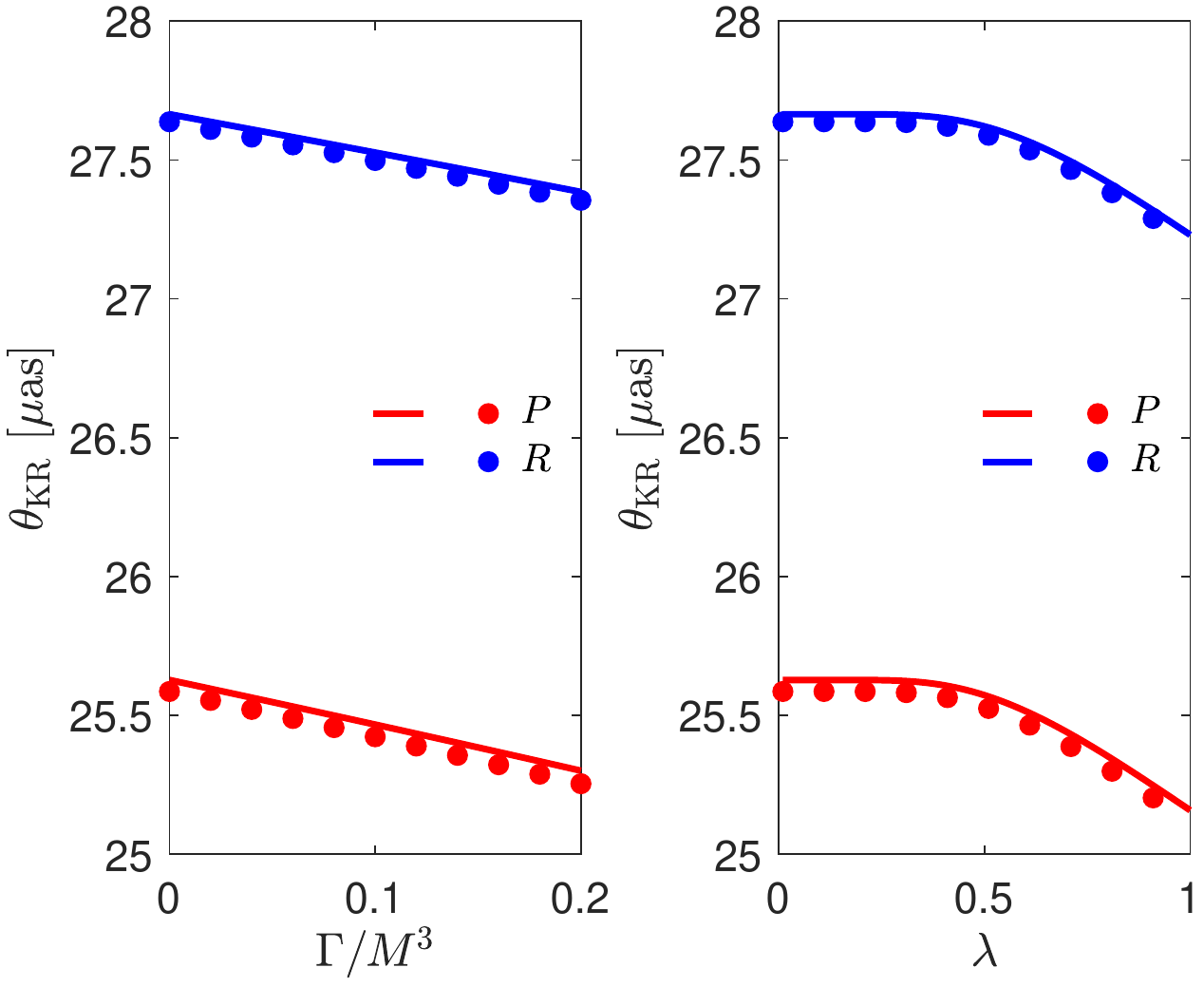}
\caption{\label{fig:krappang} The apparent angles $\theta_{n,\mathrm{KR}}$ in the KR spacetime as functions of (left) $\Gamma$ and (right) $\lambda$. The solid curves are for $n=1$ and dots are for $n=\infty$. $P$ and $R$ are for prograde and retrograde motions respectively.
We fixed $\lambda=2/3$ in (left) and  $\Gamma=\Gamma_0 M^{2/\lambda}$ with $\Gamma_0=0.1$ in (right). $\hat{a}=0.1M,~\Delta\phi_{sd}=\pi$ and other parameters are the same as in Fig. \ref{fig:kerrdphidt}. $P$ is for prograde and $R$ for retrograde motions.
}
\end{figure}

With $C_{\mathrm{KR},0}$ and $D_{\mathrm{KR},0}$ known, using Eqs. \eqref{eq:bnsol} and \eqref{eq:thetasol}, we can solve for the apparent angles $\theta_{n,\mathrm{KR}}$ of the GLed images in the SDL. Again, the effect of spin $\hat{a}$ on them  is similar to that in the Kerr case, so we only plot the dependence of $\theta_{n,\mathrm{KR}}$
on $\Gamma$ and $\lambda$ in Fig. \ref{fig:krappang}. As known from Fig. \ref{fig:krdphionother} (a) (left) that increasing $\Gamma$ will decrease $b_c$ for signals from both rotation directions, though the prograde signal decreases slightly faster than  the retrograde signal. Since the apparent angles are roughly determined by $b_c/r_d$, we also expect all apparent angles to decrease, which is exactly what we observe in Fig. \ref{fig:krappang} (left). As for the effect of $\lambda$, it is seen that the increase of $\lambda$, i.e., the transition from Kerr to KN spacetime, actually decreases the apparent angles for signals from both sides. Again, qualitatively, this is similar to the effect of charge in Reissner-Nordstr\"om spacetime, where an increase of charge also decreases the apparent angle in the SDL \cite{Pang:2018jpm}. Overall, the effects of both $\Gamma$ and $\lambda$ over reasonably large ranges are much smaller (about 0.5 [$\mu$as]) than that of the spacetime spin $\hat{a}$ (about 5.5 [$\mu$as]), as seen in Fig. \ref{fig:kerrappang}.

\begin{figure}[htp!]
\centering
\includegraphics[width=0.45\textwidth]{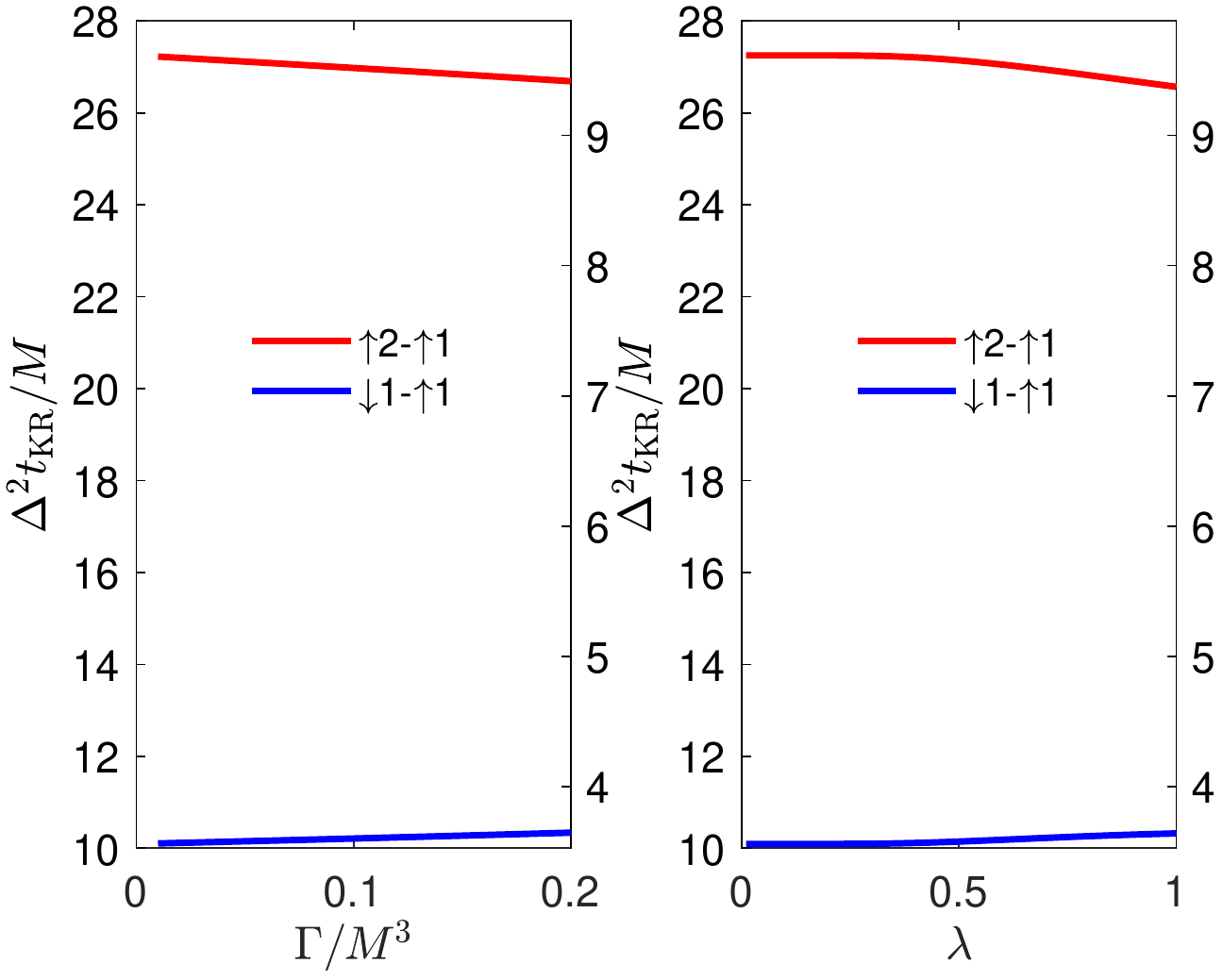}
\caption{\label{fig:krtdplot} The time delays $\Delta^2t_{\mathrm{KR}\uparrow 1,\uparrow 2}$ and $\Delta^2_{\mathrm{KR}\uparrow 1,\downarrow 1}$ in the rotating KR spacetime as functions of (left) $\Gamma$ and (right) $\lambda$. We fixed $\lambda=2/3$ in the left panel and  $\Gamma=\Gamma_0 M^{2/\lambda}$ with $\Gamma_0=0.1$ in the right panel. $\Delta\phi_{sd}=\pi$ and other parameters are the same as in Fig. \ref{fig:kerrdphidt}.
}
\end{figure}

The time delay between different signals in this spacetime is shown in Fig. \ref{fig:krtdplot}. The spin $\hat{a}$ has a similar effect on these time delays as in Kerr spacetime and therefore are not plotted. For the signals from the same side of the lens, then the effect of both $\Gamma$ and $\lambda$ can be anticipated from their effects on $r_c$ or $b_c$, which in turn can be seen from  their influence on the apparent angles in Fig. \ref{fig:krappang}. We see that as both $\Gamma$ and $\lambda$ increase, the time delays also decrease by roughly the same percentage as the decrease of $\theta_{\mathrm{KR}}$ in Fig. \ref{fig:krappang}. For the signals from different sides but with the same winding number, their time delays (retrograde flight time minus prograde flight time) slightly increases as either $\Gamma$ or $\lambda$ increase. This is consistent with the observation from Fig. \ref{fig:krappang} that as these parameters  decrease, the retrograde signals' apparent angle, and therefore critical radius and time for one cycle, decreases slightly slower than the prograde ones. Comparing this to the effect of spin $\hat{a}$ in Fig. \ref{fig:kerrtdplot}, it is seen that both $\Gamma$ and $\lambda$ affect the time delay more weakly, as is the case of their effects on the apparent angles in Fig. \ref{fig:krappang}.

\section{Summary and discussions\label{sec:disc}}

In this work, we employ a perturbative method to investigate the deflection and GL of signals with arbitrary velocity in the equatorial plane of arbitrary SAS
spacetimes. We showed that the deflection angle and total travel time can be expressed as quasi-power series of the form in Eqs. \eqref{eq:sedphi} and \eqref{eq:sedt}, i.e., $\sum_{n=0}\lsb C_n\ln \lb 1-b_c/b\rb +D_n\rsb \lb 1-b_c/b\rb^n$  in the SDL. In our approach, the coefficients $C_n,~D_n$ naturally incorporate the effects of the spacetime parameters, kinetic parameters of the signal and more importantly the finite distance effect of the source and detector. Using this deflection, we are able to establish a more exact GL equation \eqref{eq:exactgleq} from which the apparent angles of the relativistic images and time delays between them are solved as in Eqs. \eqref{eq:thetasol}, \eqref{eq:tdsamesidegeneral} and \eqref{eq:tdoppsidegeneral} for arbitrary source angular position. Applying these results to the Kerr spacetime, in Figs. \ref{fig:kerrappang}-\ref{fig:kerrtdplot} we were able to show that increasing the spacetime spin or decreasing the source angle will decrease (or increase) the apparent angles of prograde (or retrograde) signal, while decreasing the signal velocity will increase the apparent angle for all motion directions. Furthermore, the time delay between prograde (or retrograde) signals decreases (or increases), while the time delay of  a retrograde signal against prograde one increases as the spacetime spin increases. Besides, the source location also affects this time delay roughly linearly.
Finally, in the rotating KR spacetime, the increases of the effective charge parameter $\Gamma$ and the parameter $\lambda$ are found in Figs. \ref{fig:krdphionother}-\ref{fig:krtdplot} to decrease both the deflection and apparent angles of both prograde and retrograde signals and the time delay from the same rotation direction, and to increase the time delay from different sides of the lens.

A few comments about the work  are in order. Firstly, it is that the perturbative method in general seems to have a broad applicability. It can be used to signals with arbitrary velocity, in the equatorial plane of arbitrary SAS spacetime, and with arbitrary source and detector distances. The second is that the  previously known SDL deflection angle $\Delta\phi\approx C\ln (1-b_c/b)+D$ is now shown to be the leading order of a more general series. And moreover, the coefficients $C_n$ and $D_n$ in  the current work contain richer information, including the signal velocity and finite distance effect. The third is regarding the GL equation built upon the deflection angle with finite source and detector distance effect taken into account. Previously GL equations in the SDL have most often employed a deflection angle obtained for infinite source and detector distance, and many \cite{Hsieh:2021scb,Kuang:2022xjp,Islam:2021dyk,Islam:2021ful,Hsieh:2021rru} if not all are only applicable to the $\Delta\phi_{sd}\approx \pi$, i.e., the source-lens-detector nearly aligned case. However the new equation \eqref{eq:exactgleq} is more natural in that it is merely a re-statement of the definition of the deflection angle, and it can work for the case that $\Delta\phi_{sd}$ is far from $\pi$ and therefore the effect of $\Delta\phi_{sd}$ on various observables can be studied. 

Regarding the potential extension of this work, there are a few directions worth mentioning. The first and simplest, is to apply the method in this study to other SAS spacetimes, to study the effect of various parameters. The second is that we hope to extend the perturbative method to non-equatorial motion in either the weak deflection limit or the SDL. We expect this to be significantly more challenging yet also very valuable. The third is then to generalize the perturbative method to other types of motion, such as the bounded motion of much slower test particles. We believe that the perturbative method could also provide useful insights for such motions as well.

\acknowledgements

The authors wish to thank Haotian Liu for helping with the illustration in Fig. \ref{fig:schm}. This work is partially supported by the China-Ukraine IGSCP-12. The work of S. Lin is partially supported by the Undergraduate Training Programs for Innovation and Entrepreneurship of Wuhan University.

\appendix
\section{Integration formulas\label{appendix:integrals}}
In this appendix, we show how to compute the integrals of type \eqref{eq:dphitrans} and \eqref{eq:dttrans} after they are expanded by Eqs. \eqref{eq:yphindef} and \eqref{eq:ytndef}. In general, all of these integrals can be cast into the following form
\begin{align}
    I_n
    =\int_a^{\eta_i}\frac{\xi^{n/2}}{\sqrt{\xi-a}(\sqrt{\xi}+\zeta_h)}\dd\xi,~n=-1,~0,~1,~\cdots,
\end{align}
where $i=s,d$. After making a change of variable $\xi=\zeta^2$, it becomes
\begin{align}
    I_n=2\int_{\sqrt{a}}^{\sqrt{\eta_i}}{\frac{\zeta^{n+1}}{\sqrt{\zeta^2-a}(\zeta+\zeta_h)}\dd\zeta}.
\end{align}

The following steps for processing this integration then depend on whether $n=-1$ or $n\in\mathbb{Z}_\geq$. For  $n=-1$, after making a change of variable $\zeta=\sqrt{a}\cosh{s}$, it can be worked out as
\begin{align}
    I_{-1}
    =&\frac{4}{\sqrt{\zeta_h^2-a}}\mathrm{artanh}\left(\sqrt{\frac{(\zeta_h-\sqrt{a})(\sqrt{\eta_i}-\sqrt{a})}{(\zeta_h+\sqrt{a})(\sqrt{\eta_i}+\sqrt{a})}}\right).
\end{align}
To the $\mathcal{O}(a)^1$ order, this becomes
\begin{align}
I_{-1}=-\frac{1}{\zeta_h}\ln a+\frac{2}{\zeta_h}\ln{\left(\frac{2\zeta_h\sqrt{\eta_i}}{\zeta_h+\sqrt{\eta_i}}\right)}+\mathcal{O}(a)^1. \label{eq:imoneexp}
\end{align}
For $n\in\mathbb{Z}_\geq$, showing $I_n$ are also integrable although is technically possible, it is too tedious. What is needed is essentially the small $a$ limit of these integrals. In this limit, to the order $\mathcal{O}(a)^0$, the integrals can be transformed and then worked out as
\begin{align}
I_n=&2\int_0^{\sqrt{\eta_i}}{\frac{\zeta^{n}}{(\zeta+\zeta_h)}\dd\zeta}+\mathcal{O}(a)^1\nn\\
    =&2\left[\sum_{j=1}^nC_n^j\frac{(-\zeta_h)^{n-j}}{j}\left[(\sqrt{\eta_i}+\zeta_h)^j-\zeta_h^j\right]\right.\nn\\
    &\left.+(-\zeta_h)^n\ln{\left(\frac{\sqrt{\eta_i}+\zeta_h}{\zeta_h}\right)}\right]+\mathcal{O}(a)^1.
    \label{eq:ilargernexp}
\end{align}

Substituting Eqs. \eqref{eq:imoneexp} and \eqref{eq:ilargernexp} into Eqs. \eqref{eq:dphitransexp} and \eqref{eq:dttransexp}, we will be able to obtain the results \eqref{eq:sedphi} and \eqref{eq:sedt}.

\section{$p(x)$ and $q(x)$ for general metrics\label{sec:appd2}}
In this appendix, we will present the explicit series forms of the functions $p(x)$ and $q(x)$ for the metric functions \eqref{eq:gbyrc}. They are used to obtain the series expansion of the integrands for $\Delta\phi$ and $\Delta t$.

First we expand $p(x)$ defined in Eq. \eqref{eq:pp} around $1/r_c$. It can be shown that $p(1/r_c)=0$ and $p'(1/r_c)=0$, and therefore the expansion has the following form
\begin{align}
    p(x)=p_2\left(x-\frac{1}{r_c}\right)^2+p_3\left(x-\frac{1}{r_c}\right)^3+\cdots,
    \label{eq:pxseries}
\end{align}
where
\begin{align}
    p_n=\left.\frac{1}{n!}\frac{\dd^n p(x)}{\dd x^n}\right|_{x=1/r_c}.
\end{align}
Using the Lagrange inversion theorem, we can obtain the series form of $q(x)$ immediately
\begin{align}
    q(x)=\sum_{n=0}^\infty{q_n x^{n/2}},
    \label{eq:qxseries}
\end{align}
with the first few coefficients
\begin{align}
    q_0=\frac{1}{r_c},~
    q_1=-\frac{1}{\sqrt{p_2}},~
    q_2=-\frac{p_3}{2p_2^2}.
    \label{eq:q1q2}
\end{align}
The  higher-order coefficients are also straightforward to compute but they are too lengthy to be presented here.

\end{document}